\documentclass[epj]{svjour}
\usepackage{graphics}
\usepackage{epsfig}
\usepackage{amsmath}
\usepackage{cite}
\begin{document}
\title{Comprehensive measurements of cross sections and spin observables of the three-body break-up channel in deuteron-deuteron scattering at 65~MeV/nucleon}
\subtitle{}

\author{R.~Ramazani-Sharifabadi\inst{1,2}\and 
        H.R.~Amir-Ahmadi\inst{1}\and
        M.T.~Bayat\inst{1}\and
        A.~Deltuva\inst{3}\and
        M.~Eslami-Kalantari\inst{4}\and
        N.~Kalantar-Nayestanaki\inst{1}\and
        St.~Kistryn\inst{5}\and
        A.~Kozela\inst{6}\and
        M.~Mahjour-Shafiei\inst{2}\and 
        H.~Mardanpour\inst{1}\and 
        J.G.~Messchendorp \inst{1}\and
        M.~Mohammadi-Dadkan\inst{1,7}\and
        A.~Ramazani-Moghaddam-Arani\inst{8}\and
        E.~Stephan\inst{9}\and
        H.~Tavakoli-Zaniani\inst{1,4}
}                     

\institute{KVI-CART, University of Groningen, Groningen, The Netherlands\and 
 Department of Physics, University of Tehran, Tehran, Iran\and 
Institute of Theoretical Physics and Astronomy, Vilnius University, Vilnius, Lithuania\and 
Department of Physics, School of Science, Yazd University, Yazd, Iran\and
Institute of Physics, Jagiellonian University, Krak\'ow, Poland\and 
Institute of Nuclear Physics, PAS, Krak\'ow, Poland\and 
Department of Physics, University of Sistan and Baluchestan, Zahedan, Iran\and 
Department of Physics, Faculty of Science, University of Kashan, Kashan, Iran\and 
Institute of Physics, University of Silesia, Chorz\'ow, Poland
}
\date{Received: date / Revised version: date}
%
\abstract{
Detailed measurements of five-fold differential cross sections and a rich set of vector and tensor analyzing powers of the $^{2}{\rm H}(\vec d,dp){n}$ break-up process using polarized deuteron-beam energy of 65~MeV/nucleon with a liquid-deuterium target are presented. The experiment was conducted at the AGOR facility at KVI using the BINA 4$\pi$-detection system. We discuss the analysis procedure including a thorough study of the systematic uncertainties. The results can be used to examine upcoming state-of-the-art calculations in the four-nucleon scattering domain, and will, thereby, provide further insights into the dynamics of three- and four-nucleon forces in few-nucleon systems. The results of coplanar configurations are compared with the results of recent theoretical calculations based on the Single-Scattering Approximation (SSA). Through these comparisons, the validity of SSA approximation is investigated in the Quasi-Free (QF) region.
%
} 
\maketitle
\section{Introduction}
\label{intro}
\sloppy
Understanding the degrees of freedom involved in the nuclear forces is of paramount importance in subatomic physics. According to the standard model of particle physics, the nuclear force is considered to be the residual of strong interactions between quarks and gluons. It is common to interpret the interactions between nucleons by meson-exchange theory which was introduced by Yukawa in 1935. This theory successfully described the interaction between two nucleons with the exchange of virtual mesons between them~{\cite{Yukaw}}. The discovery of the pion and subsequently heavier mesons stimulated researchers to develop boson-exchange models to describe nucleon-nucleon interactions. To date, several phenomenological nucleon-nucleon (NN) potentials have been derived based on Yukawa's model~\cite{Epel1}. Some of them are successfully linked to the underlying fundamental theory of the quantum chromodynamics (QCD) by chiral perturbation theory, $\chi$PT~\cite{Epel1,Epel2}.\\
\sloppy
\indent Applying high-precision NN potentials to describe systems composed of at least three nucleons shows striking discrepancies between theoretical calculations and few particular experimental observables, despite its major successes. For instance, rigorous Faddeev calculations based on these NN potentials for the binding energy of triton, which is the simplest three-nucleon system, underestimate the experimental data~{\cite{wir}} by 10\%. In addition, they show large discrepancies with cross section data in elastic nucleon-deuteron scattering~\cite{sakai00}. These observations show that calculations based on NN potentials are not sufficient to describe systems that involve more than two nucleons. These discrepancies are not fully explained by relativistic effects since these effects are generally too small to resolve the discrepancy in all experiments even at intermediate energies~\cite{wit0,wita9}. This has led to the notion of the three-nucleon force (3NF)~{\cite{prima}}. Green's function Monte Carlo calculations based on the AV18 NN potential complemented with the IL7 three-nucleon potential give a reasonable description of the experimental data for the binding energies of light nuclei~{\cite{piep}}. Also, the inclusion of 3NF effects can partly resolve the deficiencies related to the differential cross section~{\cite{wita}}. Based on predictions of chiral perturbation theory, there is a hierarchy for few nucleon forces in a way that the 2NF is stronger than 3NF, 3NF is stronger than 4NF, and so on~\cite{Epel}. Therefore, 4NF also play a role in the four-body systems albeit a smaller one than 3NF~\cite{nasser1}.\\
\sloppy
\indent  Presently, there is an extensive database in nucleon-deuteron scattering at different energies below the pion-production threshold in elastic~{\cite{kars01,kars03,kars05,kimiko05,postma,hamid,kurodo,mermod,Igo,ald,Hos,Ela07,shimi,hatan,IUCF,Ahmad1,Ahmad2}} and break-up~{\cite{st1,st2,st3,hos2,steph,Ic4,Adl,Par,Haj,Dad,Ahmad4}} channels. 3NF effects are generally small in 3N systems except in limited parts of the phase space. For instance, the contribution of the 3NF in the cross section is expected to be relatively larger where the differential cross section of the  N$d$ elastic channel is as its minimum~{\cite{wita,nemoto}}.
 An alternative, which is the focus of this paper, is to investigate a four-nucleon (4N) system in which 3NF effects could be significantly enhanced~{\cite{nasser1}}. Deuteron-deuteron scattering, as a 4N system, is a rich laboratory to study 3NF effects in detail because of its variety of final states, observables, and kinematical configurations.\\
\sloppy
\indent Compared to the 3N systems, there is a limited experimental database for 4N systems in the low-energy regime below the three- and four-body break-up threshold~{\cite{phill,vivi,fish}}. At these low energies, the calculations are very reliable, but the effect of 3NF is very small and hard to observe. Above the break-up thresholds, the 4N database becomes even more scarce~{\cite{bech,Aldr,Garc,Micher,Ic1,Ic2,Ic3,myp2}}. Rigorous theoretical calculations for four nucleon systems are limited to energies below 30~MeV~\cite{Det1,Det2,Det3,Det4,Det5,Det6,Det7,Det8}. So far, there is no {\it ab-initio} calculation at intermediate energies. For this region, the calculations are based on some approximations~\cite{Delt1,Delt2,Delt3} which give a reasonable estimation for quasi-free scattering process. A recently dedicated research in the quasi-free domain indicates that not only the momentum of the spectator neutron should be constrained to low values, but also the momentum transfer between the beam projectile and the ejectile in the reaction should be considered~\cite{myp1}.\\
\indent In this paper, we present the results of an investigation of the $^{2}{\rm H}(\vec d,dp){n}$ break-up scattering process at a deuteron-beam energy of 65~MeV/nucleon. The data were obtained using vector- and tensor-polarized deuteron beams that were provided by the AGOR facility at KVI in Groningen, the Netherlands. Measurements of the differential cross sections and a rich set of spin observables for a large portion of the phase space were performed. This work extends the results from an earlier study reported in Ref.~\cite{Ahmad4}. The preliminary results presented in Ref.~\cite{Ahmad40} underestimated the differential cross section by a factor 2000 compared to predictions based on quasi-free approximation. In the analysis presented here, we identified a normalization error in the earlier study. The cross section data are, therefore, corrected according to our new insights. The results of coplanar configurations are compared with the results of recent theoretical calculations based on Single-Scattering Approximation~\cite{Delt1,Delt2,Delt3}.  The results presented here are the most precise and accurate data of the $^{2}{\rm H}(\vec d,dp){n}$ process, identified from many hadronic reaction channels, at 65 MeV/nucleon. The aim is to provide high-precision data to study in detail the role of 3NF in few-body systems once {\it ab-initio} calculations become available.
\section{Theoretical approach}
\label{sec2}
\sloppy
Exact {\it ab-initio} description of four-nucleon scattering has been successfully accomplished
for two-cluster reactions up to about 30 MeV energy \cite{Det1,Det2,Det3,Det4,Det5,Det6,Det7,Det8,Delt1,Delt2},
but break-up reactions at intermediate energies are beyond the present developments.
Instead, an approximate treatment of the three-body break-up in the deuteron-deuteron
collisions was proposed in Ref.~\cite{Delt3}. It was based on the first term of the Neumann series
expansion of exact Faddeev-Yakubovsky equations for  transition operators in the
momentum space, where the three-cluster break-up operator 
\begin{equation} \label{eq:U3}
\mathcal{U}_{32} = (1- P_{34}) U_1,
\end{equation}
is approximated by the three-nucleon transition operator $U_1$ with
the permutation operator $P_{34}$.
This approximation corresponds to the full interaction between the deuteron beam and 
one nucleon in the deuteron target, while the remaining nucleon experiences no interaction
with the deuteron beam. As a consequence, the interacting nucleon is knocked out
with a significant momentum, while the second one acts as a spectator. The momentum distribution of the spectator is given by the deuteron bound state wave function, implying
that momentum and energy of this deuteron remain predominantly low,
the condition of quasi-free scattering.
If the relative energy  between this spectator nucleon and two other outgoing clusters 
is high enough such that their interactions can be neglected, this may become a reasonable
approximation under the quasi-free kinematical conditions.
The above approximation for the proton knockout from the target deuteron is labeled as
SSA1. The amplitude in Eq.~\ref{eq:U3} taken between properly symmetrized initial
and final states has four contributions, corresponding to the permutations
of beam and target deuterons and of two nucleons within the broken deuteron;
the corresponding result is labeled as SSA4. Note that under the quasi-free scattering
conditions, the SSA1 term is dominant, such that the remaining three terms in SSA4 are 
insignificant and the full SSA4 result is close to the SSA1 one.

The details of the SSA1 and SSA4 approximations have been reported in Ref.~\cite{Delt3} with several
realistic nucleon interaction models, revealing a small sensitivity to the 
force model. For this reason, the present study uses only one potential, 
the CD Bonn + $\Delta$ \cite{Delt1},
that explicitly includes a virtual excitation of a nucleon to a $\Delta$ isobar, thereby, providing an effective 3NF.
\section{Experimental setup}
\label{Exp}
\indent The experimental setup used to investigate the deuteron-deuteron scattering process is BINA,  \textbf{B}ig-\textbf{I}nstrument for \textbf{N}uclear-Polarization \textbf{A}nalysis. Unpolarized and vector- and tensor-polarized deuteron beams were produced by the atomic Polarized Ion Source (POLIS) with nominal polarization values between 60-80\% of the maximum theoretical values~\cite{Fri,Krem}. The beam was accelerated to 130 MeV by the superconducting AGOR facility. The accelerated deuteron beam impinged on the liquid-deuterium target inside the scattering chamber of BINA during a measurement period of about 51 hours~{\cite{nasser3}. The thickness of the target cell was 3.85~mm with an uncertainty of 5\%. BINA was developed with nearly full coverage of the geometrical acceptance and it is capable of measuring the energy and scattering angles of the outgoing charged particles with high-resolution especially in the forward part, and it provides information for particle identification~{\cite{nasser,hos3,Ahmad3}}.\\
\hspace{-1cm}\begin{figure}[t]
 \centering 
\epsfxsize=8.5 cm
\epsfbox{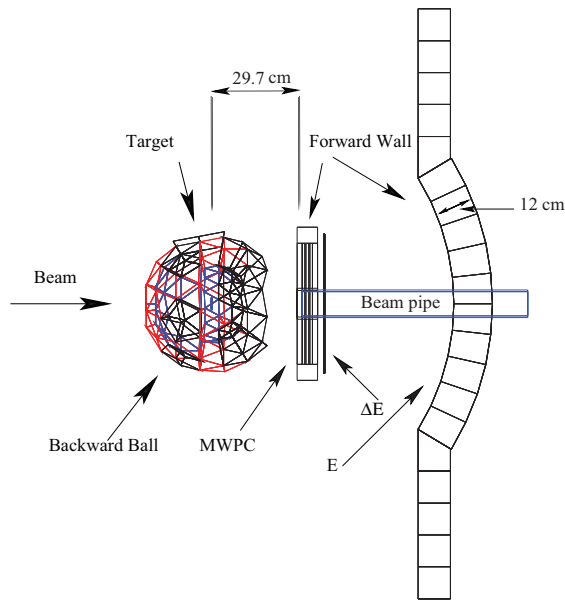} 
\vspace*{0.0cm}
\caption{A side view of BINA. The elements on the right side show of the forward part of BINA, including the
  multi-wire proportional chamber (MWPC), an array of twelve thin plastic ($\Delta E$) scintillators followed by ten thick segmented ($E$) scintillators
  mounted in a cylindrical shape.
  On the left side, the backward part of BINA is depicted, composed of 149 phoswich scintillators glued together to form the scattering chamber.}\label{bin}
\end{figure}
\indent Figure~\ref{bin} shows a sketch of BINA. The setup consists of two parts, a forward wall and a backward ball.
The forward wall consists of a multi-wire proportional chamber (MWPC) to determine the scattering angles of the particles,
twelve vertically-mounted plastic $\Delta E$ scintillators with a thickness of 2~mm, and ten horizontally-mounted $E$ scintillators with a thickness of 12~cm.
The $E$ scintillators are mounted in a cylindrical shape with the center of the cylinder matching the interaction point of the beam with the target. Although, the $\Delta E$-$E$ hodoscope provides the possibility to
perform particle identification, the information from the $\Delta E$ detectors was not used in this experiment for particle identification, because the yield of scintillation light reaching the photomultipliers was not sufficient for this purpose. During a visual inspection after the experiment, these scintillators were found to be damaged.
 The particle identification was done using the time-of-flight (TOF) information of the $E$ scintillators. Photomultiplier tubes (PMTs) were mounted on both sides of each $E$ scintillator. Signals from these PMTs are used to extract the energy and TOF of the scattered particles. The resolution of TOF is around 0.5 ns.\\
\indent The MWPC covers scattering angles between 10$^\circ$ and 32$^\circ$ with a full azimuthal angle coverage and up to 37$^\circ$ with a limited azimuthal angle coverage. The MWPC has a resolution of $0.4^{\circ}$ for the polar angle and between $0.6^{\circ}$ and $2.0^{\circ}$ for the azimuthal angle depending on the polar
scattering angle. The scattering angles, energies and TOF of the final-state deuterons and protons were measured by the MWPC and the scintillators of BINA. The backward ball of BINA is made of 149 phoswich scintillators that were used as detector and scattering chamber with a scattering-angle coverage between $40^\circ$ and $165^\circ$ and nearly full azimuthal coverage. The ball detectors were not used in the analysis presented in this paper. For a detailed description of the BINA, we refer to Ref.~\cite{hos3}.\\
\indent A Faraday cup was mounted at the end of the beam line to monitor the beam current throughout the experiment. The typical current of the deuteron beam was 4~pA. The current meter of the Faraday cup was calibrated using a current source with an uncertainty of 2\%~{\cite{Ahmad4}}. A small offset of 0.28 $\pm$ 0.13~pA in the readout of the current was observed. The offset has been determined by minimizing the reduced $\chi^2$ whereby an offset in the current is introduced as a free parameter using the comparison between the results of the $Re(T_{22})$ from the elastic channel of $dd$ scattering and those coming from the magnetic Big-Bite Spectrometer (BBS) at KVI~{\cite{BBSr}}. The error is obtained by evaluating the $\chi^2$ distribution as a function of the offset. The intersection point of this distribution with a $\chi^2$ value that is one unit larger than its minimum has been used to determine the uncertainty in the offset.\\
\indent The polarization of deuteron beam was monitored with a Lamb-Shift Polarimeter (LSP)~\cite{lsp} at the low-energy beam line and measured with the BINA after the beam acceleration using a measurement of the elastic deuteron-proton scattering process~\cite{ibp}. The vector and tensor polarizations of the deuteron beams were found to be $p_{Z}=-0.601\pm 0.029$ and $p_{ZZ}=-1.517\pm 0.032$, respectively, whereby the errors include uncertainties in the analyzing powers of the elastic deuteron-proton scattering. The polarization of the deuteron beam was monitored for different periods of experiment and found to be stable within statistical uncertainties~{\cite{Ahmad4}}.
\section{Event selection and analysis method}
\label{Ana}
\indent Differential cross sections and spin observables of the three-body break-up process have been measured in a nearly background-free experiment. The identification of the three-body break-up channel amongst other hadronic channels has been done using the measured energies, scattering angles, and TOF information of detected particles. The hardware trigger was biased on the selection of events for which two particles were registered in coincidence by the forward wall of BINA corresponding to small scattering angles.\\
\indent In this analysis, the correlation between the energies of the deuteron and the proton ($S$-curve) for a desired configuration, $(\theta _d, \theta _p, \phi_{12})$ is obtained. In total, there are nine variables involved in the three-body break-up process, namely $\theta_i$, $\phi_i$, and $E_i$ where $i$ refers to the deuteron, proton, or neutron. Considering the momentum and energy conservation laws, measuring kinematical variables of two particles is enough to obtain the other variables unambiguously. As a result, the kinematics of each configuration of the three-body break-up channel are specified by the scattering angle of the deuteron, $\theta _d$, the scattering angle of the proton, $\theta _p$, the difference between azimuthal angles of the deuteron and proton, $\phi _{12}=|\phi _d-\phi _p|$, and the correlation between the energies of the two particles, $E_d$ and $E_p$. For convenience, in the analysis, the energies of the deuteron and proton are also expressed by two new variables, namely $S$ and $D$. The variable $S$ is the arc-length along the $S$-curve with the starting point at the minimum energy of the deuteron. The variable $D$ is the distance between the point ($E_d$, $E_p$) and the kinematical $S$-curve. Figure~\ref{scurve} shows the theoretical $S$-curves of some configurations in the three-body break-up channel. For instance, the configuration (20$^\circ$, 20$^\circ$, 180$^\circ$) refers to a deuteron scattering to 20$^\circ \pm$1$^\circ$, a proton scattering to 20$^\circ \pm$1$^\circ$, and the difference between azimuthal angles of the two particles is 180$^\circ \pm$5$^\circ$. The expected energy correlation ($E_d$, $E_p$) is used to calibrate the energy of the particles in the break-up channel. Besides, it is used to count the number of signal events from a spectrum including background. This procedure will be discussed later in this section.\\
\hspace{-1cm}\begin{figure}[t]
 \centering 
\epsfxsize=9.1cm
\epsfysize=8.cm
\epsfbox{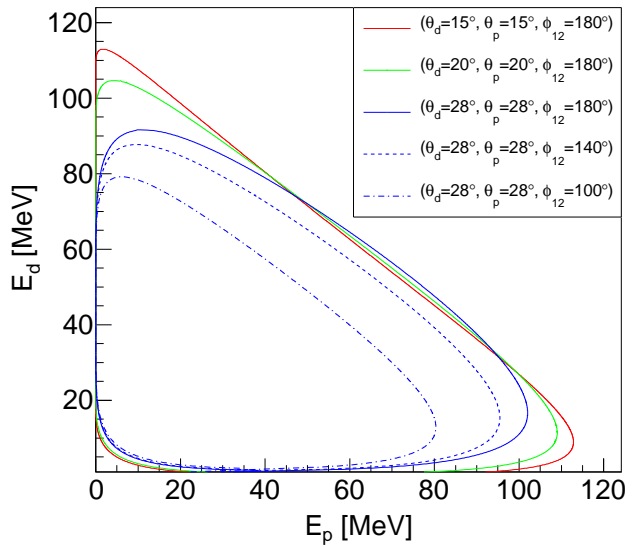} 
\vspace*{-0.35cm}
 \caption{Correlation between initial energies of deuteron and proton detected in coincidence in the forward wall of BINA, for selected configurations indicated in the inset. The kinematics of each configuration, $(\theta _d, \theta _p, \phi_{12})$, is defined by scattering angle of deuteron, $\theta_{d}$, scattering angle of proton, $\theta_{p}$, and the relative azimuthal angle of the two particles, $\phi_{12}$.}\label{scurve}
\end{figure}
\indent The scintillator response was calibrated by matching the data to the expected energy correlation between $E_p$ and $E_d$ for each configuration of the break-up channel. Two calibration methods have been exploited to investigate the sensitivity of final results to the procedures. In the first calibration method, we followed the procedure introduced in Ref.~\cite{Ahmad3}. In this method, the detector response has been parametrized by a non-linear two-parameter function. 
In the second method, a fourth-order polynomial function was used to fit the experimental data points to the theoretical $S$-curve. The average of the results obtained for cross sections and analyzing powers based on the two energy calibration procedures was used as final results for each data point; see Sec.~\ref{results}. The difference between these two results is used to estimate a systematic uncertainty for each data point attributed to calibration errors. The energy losses between the interaction point and the scintillators were accounted for via Monte Carlo studies using a model of BINA implemented in GEANT3~\cite{ge3}.\\
\indent The next step is to describe the technique that is used to identify the type of particle, namely proton or deuteron, in the $^{2}{\rm H}(\vec d,dp){n}$ reaction. Particle identification is performed by comparing the measured relative TOF of the registered particles with the value calculated on the basis of the kinematics of the three-body break-up reaction. This procedure was checked with the information of the missing mass of the undetected particle, namely the neutron, and the data, after applying particle identification. A detailed description of this check can be found in Ref.~\cite{Ahmad4}.\\
\hspace{-1cm}\begin{figure}[t]
 \centering 
\epsfxsize=9.2cm
\epsfbox{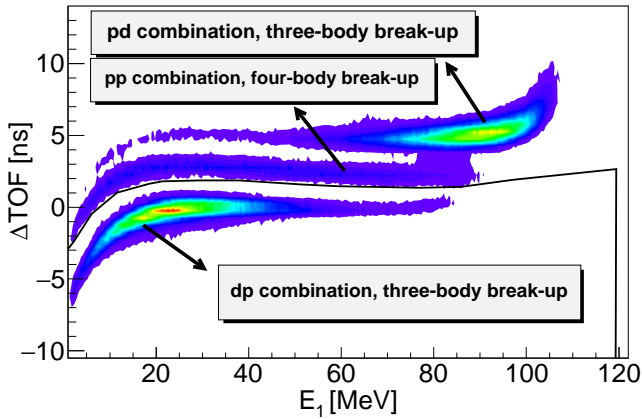} 
\vspace*{-0.5cm}
 \caption{$\Delta TOF$ as a function of the energy of the particle we assumed to be deuteron. The scattering angles of two particles are fixed to be 25$^\circ \pm 1^\circ$ with a relative azimuthal angle of 180$^\circ \pm 5^\circ$. The data concentrated in the bottom (top) band correspond to the three-body breakup channel for which the first particle is a deuteron (proton) and the second one a proton (deuteron), namely $dp$ combination ($pd$ combination). The middle one corresponds to the four-body break-up channel whereby the two detected particles are both protons ($pp$ combination). The solid line is used as a graphical cut to select the valid combination.}\label{dtof}
\end{figure}
\indent The Time-to-Digital Converter (TDC) outputs corresponding to the PMTs of the left ($TDC_L$) and right ($TDC_R$) side of each scintillator, corrected for time-walk effect by applying the ifnormation from the analog Constant-Fraction Discriminators, are added for each event. The TDCs were operating in a common-stop mode. The start signal of the TDC comes from the individual PMT signals and the stop signals come from the trigger. The sum of the left and right TDCs, $TOF_i=(TDC_L)_i+(TDC_R)_i$ gives a measure of the flight time of the particle $i$ independent on hit position of the incident particle in the scintillator bar. The index $i$ refers to the deuteron and proton hits in the scintillators in the forward wall. The difference between the obtained TOFs of the deuteron and proton from the information of the TDCs, $(TOF_d-TOF_p)_{TDC}$, and that calculated from the energies and path lengths of the particles, $(TOF_p-TOF_d)_{E}$, is calculated:
\begin{equation}
\Delta TOF=(TOF_d-TOF_p)_{TDC}-(TOF_p-TOF_d)_{E}.
\label{eq1}
\end{equation}
\\
$\Delta TOF$, is in the first order, independent of the value of $S$. Since the TDC signals are used in a common-stop mode, the TDC values are opposite compared to the calculated values in Eq.~\ref{eq1}.\\
\indent Figure~\ref{dtof} shows the $\Delta TOF$ as a function of the energy of the particle we assumed to be deuteron before particle identification and for a specific configuration. The scattering angles of the two particles are fixed to be $25^\circ \pm 1^\circ$ with a relative azimuthal angle of $180^\circ \pm 5^\circ$. There are three concentrations of data in the spectrum that are distinguishable. For the calculation of the $(TOF_p-TOF_d)_{E}$ in this spectrum, we assume that the first particle is deuteron and the second particle is proton. This assumption is validated when $\Delta TOF$ is close to zero, which is the case for the bottom band. It should be emphasized that another assumption where the first particle is proton and the second particle is deuteron is considered in the event selection as well. Also, we used the energy of the deuteron and proton at the interaction point, thereby, not taking into account energy losses while traveling to the detector. Deviation from zero in the band at the bottom (also for other bands) is due to neglected energy losses of the particles when traveling from the target to the detector.\\
\hspace{-1cm}\begin{figure}[t]
 \centering 
\epsfxsize=8.8cm
\epsfysize=7.cm
\epsfbox{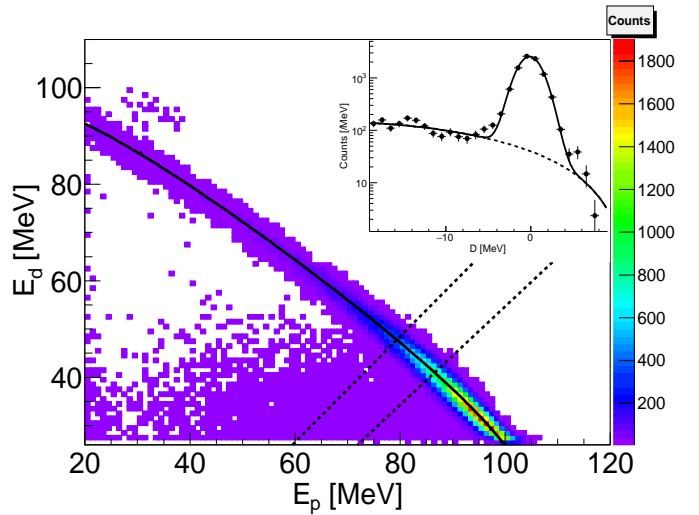} 
\vspace*{-0.5cm}
 \caption{The energy correlation between deuteron and proton for the configuration $(\theta _d, \theta _p, \phi_{12})$=($25^\circ \pm 1^\circ$, $25^\circ \pm 1^\circ$, $180^\circ \pm 5^\circ$) after particle identification. The solid line represents the $S$-curve resulting from the kinematical calculation. The dashed lines indicate the choice of one of the $S$-interval. The inset figure shows the result of projected events on the axis perpendicular to the S-curve for this $S$-interval.}\label{scurv}
\end{figure}
\indent Figure~\ref{scurv} shows the energy correlation between the deuteron and proton for the configuration $(\theta _d, \theta _p, \phi_{12})=(25^\circ \pm 2^\circ, 25^\circ \pm 2^\circ, 180^\circ \pm 5^\circ)$ after particle identification as described above. The tail on the left-hand side of the $S$-curve is dominated by events which have undergone hadronic interactions. The solid line represents the expected kinematical $S$-curve. The $S$-curve is divided into $S$-bins with a width $\Delta S$ of 10 MeV. One of those bins is indicated in the figure by the dashed lines. To count the number of break-up events in the interval of $S-\frac{\Delta S}{2}$ and $S+\frac{\Delta S}{2}$, the events are projected onto the $D$-axis perpendicular to the $S$-curve for each $S$-bin. The result of projected events for a particular $S$-bin is presented in the inset of Fig.~\ref{scurv}. This spectrum consists of a peak around zero which belongs to the break-up events with a negligible amount of accidental background, as can be seen from the small number of events on the right-hand side of the peak. The peak shows that in the majority of the break-up events, particles deposit all their energies in the scintillator. The tail on the left-hand side of the peak corresponds to events which have undergone hadronic interactions inside the scintillator. The solid line in the inset of Fig.~\ref{scurv} is the result of a fit through the data based on a Gaussian-distributed signal combined with a third-order polynomial representing the hadronic interactions contribution. Cross sections are obtained by counting the number of events under the peak, thereby subtracting the small background contribution. For the extraction of the analyzing powers, we simply counted the total number of events within a window in $D$ of $\pm$6~MeV. The fraction of break-up events for which hadronic interactions occur, and, thereby, not seen as signal events, is estimated by a GEANT3 simulation. The data are corrected for the loss of break-up events due to hadronic interactions. The average loss due to hadronic interactions for the deuteron (proton) is 16$\%\pm2\%$ (10$\%\pm2\%$)~\cite{Ahmad4}.\\
\indent  To determine the cross section for each configuration, the extracted number of counts is corrected by efficiencies of the system such as the live-time of the data acquisition, MWPC efficiencies, losses due to hadronic interactions, and down-scale factor that was used in the hardware trigger. The average live-time of the data acquisition of BINA is around $40\%$. The MWPC efficiencies for the deuteron and proton are 99$\%\pm1\%$ and 97$\%\pm1\%$, respectively~\cite{Ahmad4}. \\
\indent For the extraction of the differential cross section, one of the main uncertainties of the system comes from the thickness of the liquid-deuterium target due to the bulging effect which is around $5\%$. Other systematic uncertainties originate from the error in the efficiency of the MWPC for the deuteron (proton) of $1\%$ ($3\%$), and errors in determining the amount of hadronic losses for the deuterons and protons which are estimated to be $2\%$. Furthermore, a systematic uncertainty is considered due to the calibration procedure that is dominant in the extraction of the differential cross section. This uncertainty is included for each data point of the measured differential cross section and varies between 5$\%$ and 20$\%$. An additional systematic error stems from the uncertainties in the determination of the beam luminosity. The luminosity was continuously monitored by a Faraday cup mounted at the end of the beam line. The readout of the Faraday cup was also seen to have a small offset of 0.28$\pm$ 0.13 pA. The effect of all the uncertainties in the luminosity measurement on the extraction of the differential cross section amounted to an uncertainty of around 7$\%$. A detailed discussion on the various systematic uncertainties can be found in Ref.~\cite{mythesis}.\\
\indent The interaction of a polarized beam with an unpolarized target produces an azimuthal asymmetry in the cross section. This asymmetry is proportional to the product of the magnitude of polarization and another observable, namely analyzing power. Vector- and tensor-polarized beams give the possibility to measure various analyzing powers by studying the azimuthal variations in the cross section. The cross section of any reaction with a polarized beam is defined as~\cite{Ohl}:
\begin{multline}
\sigma(\xi,\phi)=\sigma_{0}(\xi)[1+\sqrt{3} p_{Z}Re(iT_{11}(\xi))\cos\phi\\
+\sqrt{3} p_{Z}Im(iT_{11}(\xi))\sin\phi-\frac{1}{\sqrt8}p_{ZZ}T_{20}(\xi)\\
-\frac{\sqrt3}{2} p_{ZZ}Re(T_{22}(\xi))\cos2\phi-\frac{\sqrt3}{2} p_{ZZ}Im(T_{22}(\xi))\sin2\phi],
\label{eq2}
\end{multline}
where $\sigma$ ($\sigma_0$) is the five-fold differential cross section of the reaction with polarized (unpolarized) beam and $\xi$ represents the kinematical variables of each configuration, $(\theta _d, \theta _p, \phi_{12}, S)$. $p_Z$ and $p_{ZZ}$ are the vector and tensor polarizations, respectively. $Re(iT_{11})$ and $Im(iT_{11})$ ($Re(T_{20})$, $Re(T_{22})$, and $Im(T_{22})$) are vector (tensor) analyzing powers and $\phi$ is azimuthal scattering angle of the deuteron. In this experiment, the quantization axis ($Y$-axis) is perpendicular to the beam direction ($Z$-axis) and $\phi$ is the azimuthal angle of the outgoing particle with respect to the corresponding $X$-axis. All five analyzing powers were extracted by analyzing the three-body break-up channel.\\
\indent To extract all sets of analyzing powers, we define a new function:
\begin{equation}
f^{\tilde{\xi}, \phi_{12}}(\phi)= \frac{\sigma(\xi,\phi)}{\sigma_{0}(\xi)},
\label{eq3}
\end{equation}
where $\xi \equiv (\tilde{\xi}, \phi_{12})$. Here, $\tilde{\xi}$ defines all kinematical variables excluding $\phi_{12}\equiv \phi_{d}-\phi_{p}$. Mirror configurations are those kinematical configurations  that differ only in the sign of the relative azimuthal angle $\phi_{12}$. The imaginary parts of analyzing powers in Eq.~\ref{eq2} are odd under the parity conservation while the other three analyzing powers are even. The mirror configurations $(\tilde{\xi}, \phi_{12})$ and $(\tilde{\xi},-\phi_{12})$ are used to construct two combinations of asymmetries:
\begin{equation}
g^{\xi}(\phi)=\frac{f^{\tilde{\xi}, \phi_{12}}(\phi)+f^{\tilde{\xi},-\phi_{12}}(\phi)}{2},
\label{eq4}
\end{equation}
and
\begin{equation}
h^{\xi}(\phi)=\frac{f^{\tilde{\xi}, \phi_{12}}(\phi)-f^{\tilde{\xi},-\phi_{12}}(\phi)}{2},
\label{eq5}
\end{equation}
where $g^{\xi}(\phi)$ and $h^{\xi}(\phi)$ are:
\begin{multline}
g^{\xi}(\phi)=1+\sqrt{3} p_{Z}Re(iT_{11}(\xi))\cos(\phi)\\
-\frac{1}{\sqrt8}p_{ZZ}T_{20}(\xi)-\frac{\sqrt3}{2} p_{ZZ}Re(T_{22}(\xi))\cos(2 \phi),
\label{eq6}
\end{multline}
and
\begin{multline}
h^{\xi}(\phi)=+\sqrt{3} p_{Z}Im(iT_{11}(\xi))\sin(\phi)\\
-\frac{\sqrt3}{2} p_{ZZ}Im(T_{22}(\xi))\sin(2 \phi).
\label{eq7}
\end{multline}
\indent Using the data collected with a pure vector-polarized beam, ($p_{ZZ}=0$), the $Re(iT_{11})$ is extracted from the amplitude of $\cos\phi$ component in Eq.~\ref{eq6}. Data extracted from a pure tensor-polarized beam, ($p_{Z}=0$), produce a $\cos2\phi$-shape of the azimuthal asymmetry with an offset from one due to the term, $\frac{1}{\sqrt8}p_{ZZ}T_{20}(\xi)$ in Eq.~\ref{eq6}. The amplitude of the $\cos2\phi$-shape yields $Re(T_{22})$ and the offset (departure from one) gives $Re(T_{20})$. Figure~\ref{Asymm1} shows the data for $g^{\xi}(\phi)$ as a function of $\phi$ for a pure vector-polarized beam (top panel) and pure tensor-polarized beam (bottom panel) for the kinematical point $\xi \equiv (\theta _d, \theta _p, \phi_{12}, S)$=(25$^\circ$, 20$^\circ$, 180$^\circ$, 240~MeV).\\
\indent In the same way, using data collected with a pure vector-polarized beam, ($p_{ZZ}=0$), the $Im(iT_{11})$ can be  extracted from the amplitude of $\sin\phi$ component in Eq.~\ref{eq7}. Similarly, the analyzing power, $Im(T_{22})$,  is extracted from the amplitude of the $\sin2\phi$ of Eq.~\ref{eq7} using data taken with a pure tensor-polarized beam ($p_Z=0$). Figure~\ref{Asymm2} shows the data for $h^{\xi}(\phi)$ as a function of $\phi$ for a pure vector-polarized beam (top panel) and pure tensor-polarized beam (bottom panel) for the kinematical configuration $\xi \equiv (\theta _d, \theta _p, \phi_{12}, S)$=(25$^\circ$, 20$^\circ$, 160$^\circ$, 230~MeV).\\
\begin{figure}[t]
 \centering 
\epsfxsize=10cm
\epsfbox{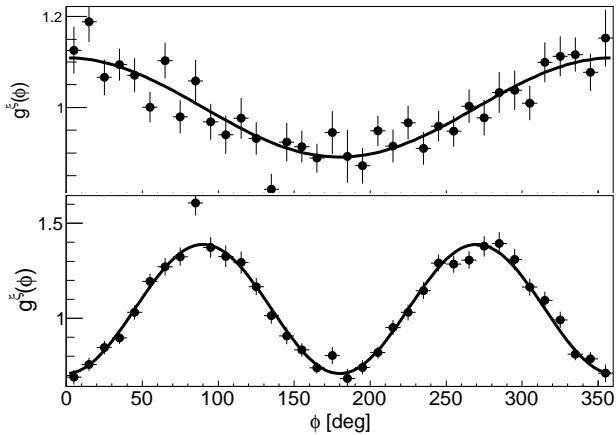} 
\hspace{-1.5cm}
\vspace*{-0.5cm}
 \caption{The data for $g^{\xi}(\phi)$ as a function of $\phi$ for a pure vector-polarized beam (top panel) and pure tensor-polarized beam (bottom panel) for the kinematical configuration $\xi \equiv (\theta _d, \theta _p, \phi_{12}, S)$=(25$^\circ$, 20$^\circ$, 180$^\circ$, 240~MeV). The $\chi^2$/35 for the top (bottom) panel is 0.97 (0.91).}\label{Asymm1}
\end{figure}
\hspace{-1.5cm}\begin{figure}[t]
 \centering 
\epsfxsize=10cm
\epsfbox{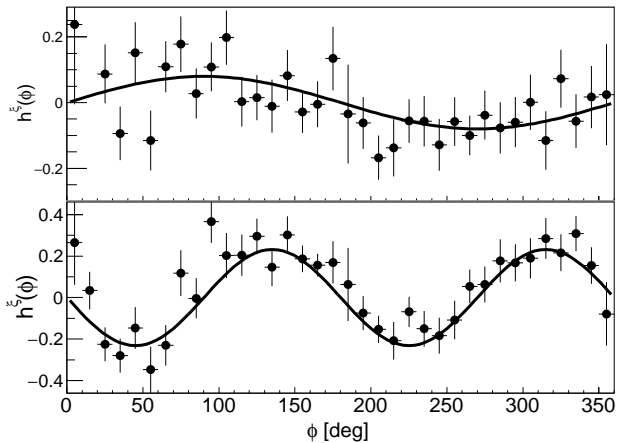} 
\vspace*{-0.5cm}
 \caption{The data for $h^{\xi}(\phi)$  as a function of $\phi$ for a pure vector-polarized beam (top panel) and pure tensor-polarized beam (bottom panel) for the kinematical configuration $\xi \equiv (\theta _d, \theta _p, \phi_{12}, S)$=(25$^\circ$, 20$^\circ$, 160$^\circ$, 230~MeV). The $\chi^2$/35 for the top (bottom) panel is 1.10 (1.14).}\label{Asymm2}
\end{figure}\\
\indent In the case of measuring analyzing powers, systematic uncertainties include the errors in measuring the polarization (4$\%$), and the beam luminosity ($2\%$). Another systematic uncertainty comes from the offset in the readout of the current. This offset imposes a shift on both polarized and unpolarized cross sections in the same direction. Such a shift does not cancel and causes an additional offset in the ratio of $\sigma/\sigma_0$, and therefore, imposes sizeable systematic uncertainties on $Re(T_{20})$. The absolute value of the average shift in the results of $Re(T_{20})$ due to the offset is around 0.05 while the measured values of this observable varies between $-0.4$ and +0.2. However, the effect of the offset in $\sigma/\sigma_0$ is very marginal on $Re(iT_{11})$ and $Re(T_{22})$, since these observables are primarily sensitive to the amplitude of the $\cos\phi$ and $\cos 2\phi$ oscillations. Finally, for the extraction of analyzing powers, the absolute value of the average uncertainty due to the calibration procedure is around 0.05 while the measured values of the spin observables varies between $-0.5$ and +0.4. Since the vector (tensor) polarization of the beam could have small impurity with a tensor (vector) component, we considered as well a fit through the data whereby we accounted for $\sin2\phi$ and $\cos2\phi$ ($\sin\phi$ and $\cos\phi$) component with amplitudes that are taken as free parameters. The resulting analyzing powers using this procedure were in a very good agreement with results that were obtained using the nominal procedure. We, therefore, did not take into account a systematic error due to impurity of the polarization. Moreover, as already mentioned, the background contribution is found to be negligible and, therefore, it has hardly any effect in the extraction of the spin observables. A detailed discussion on the various systematic uncertainties can be found in Ref.~\cite{mythesis}.\\
\section{Experimental results}
\label{results}
\indent In the present work, the differential cross sections and five analyzing powers for the three-body break-up process in the $^{2}{\rm H}(\vec d,dp)n$ reaction have been extracted as a function of $S$ for various configurations. Two energy calibration methods have been used to extract the observables. For the data presented in Figs.~\ref{Crse15}-\ref{ImT22_28}, results of the two methods are averaged. The differences between the results of the two methods are indicated by a gray band for each configuration. This error is the dominant source of the total systematic uncertainty. The contribution of the other systematic uncertainties are around 7$\%$ (5$\%$) of each measured value of the differential cross sections (analyzing powers). The kinematical variables of each configuration are indicated in each figure. For instance, the scattering angle of the deuteron is indicated in the caption of each figure. The scattering angles of the proton for each panel (configuration) are shown on the right-hand side of each figure. The absolute values of the relative azimuthal angles, $\phi_{12}$, are indicated at the top of each figure. The preliminary results of Ref.~\cite{Ahmad3} underestimated the differential cross sections by a factor of 2000 due to an error in the conversion of units and a factor two because of missing half of the statistics while performing the particle identification. These errors were identified and corrected in the re-analysis of the data in the present work.\\
\indent The solid lines in Figs.~\ref{Crse15}-\ref{T22_28} represent the results of a recent theoretical calculation based on the single-scattering approximation (SSA) by using the CD-Bonn+$\Delta$ potential for the three analyzing powers and cross sections of the coplanar configurations. As discussed in Sec.~\ref{sec2}, in SSA1, only the term related to the deuteron-target break-up with neutron acting as a spectator is included while in SSA4, all four terms are included~\cite{Delt1,Delt2,Delt3}.\\
\indent We note that the validity of SSA is limited to the quasi-free domain. Therefore, we have left out the results of the calculations for non-coplanar configurations. Very large discrepancies are indeed found between data and theory for non-coplanar cases as shown in Ref.~\cite{Ic2}. We expect, however, that the single-scattering approximation provides a reasonable estimate of the observables of the three-body break-up reaction in the quasi-free configurations. The red (black) line represents the results of SSA1 (SSA4) calculations in each panel. The blue-dashed lines indicate the energy of the neutron as the third particle in unit of MeV. We define the quasi-free scattering (QFS) region by those configurations at which the energy of the neutron is less than 300~keV corresponding to the Fermi motion of nucleons inside the deuteron. Note that a recent analysis in this region reveals that not only the momentum of neutron should be constrained to be ``in'' the QF region but also the momentum transfer between the beam projectile and the ejectile in the analysis should be taken into account~\cite{myp1}. In general, there is a good agreement between the data and the theoretical calculations in the QFS region. Note that the QFS region corresponds to the part of phase space at which the cross section peaks according to the SSA1 and SSA4 calculations. At these configurations, the results of SSA1 and SSA4 give similar predictions indicating that the SSA1 indeed term dominates in the more extended SSA4 calculations.\\
\indent Figures~\ref{Crse15}-\ref{Crse28} show that the single-scattering approximation for the differential cross sections follows the experimental data very well within the systematic uncertainties in the QFS region where the energy of the neutron is very close to zero. These regions, easy to see by following the blue-dashed line indicating the energy of the neutron, generally correspond to large scattering angles of the deuterons and the protons ($\theta_{d}>25^\circ$ and $\theta_{p}>20^\circ$). Also, comparing the experimental spin observables and the theoretical estimations reveals a general agreement, particularly in the QFS region. But a more detailed inspection of Figs.~\ref{iT11_15}-\ref{iT11_28} show that the SSA underestimates the vector analyzing power, $Re(iT_{11})$, for the configurations with smaller scattering angles of proton while it agrees perfectly with the data for the configurations with large scattering angles of the proton. This can be related to final-state interaction effects being stronger when the particles move close to each other. Also, it can be concluded that the regions for which SSA1 (red line) and SSA4 (black line) predictions are close to each other, the SSA follows the data very well. By comparing the results of the tensor analyzing powers ($Re(T_{20})$ and $Re(T_{22})$), a general agreement is observed between the results of SSA and the data in most of the configurations in the QFS region; see Figs.~\ref{T20_15}-\ref{T22_28}. We observe that also in this case, the QF regions where SSA1 (red line) and SSA4 (black line) predictions are close to each other, there is a very good agreement between SSA calculations and the data. \\
\indent The two imaginary parts of the analyzing powers, namely $Im(iT_{11})$ and $Im(T_{22})$, should be zero for coplanar configurations due to parity conservation. Our data agree well with this prediction confirming the symmetries of the scattering experiment; see Figs.~\ref{ImiT11_15}-\ref{ImT22_28}. For the non-coplanar configurations, our analysis offers a high-precision database which can be used for the future calculations including interactions between all four particles.\\
\indent In summary, we have presented a thorough analysis of the three-body break-up process of $^{2}{\rm H}(\vec d,dp)n$, at 65 MeV/nucleon. The experiments were conducted at the AGOR facility at KVI using the BINA 4$\pi$-detection system. The three-body break-up reaction has been identified by using the information of Time-of-Flight, scattering angles, and energies of the particles. We provided a rich set of measured cross sections and analyzing powers for 72 configurations. The results of coplanar configurations for the differential cross section and three analyzing powers, that are even under parity conservation, are compared with the recent theoretical calculations based on Single-Scattering Approximation. As a general conclusion, despite the fact that SSA is an approximation for a four-body interaction, the single-scattering approximation generally produces observables for a four-body interaction with respectable quality. Together with the upcoming state-of-the-art {\it ab-initio} calculations, these data will provide further insights into the dynamics of three- and four-nucleon forces in few-nucleon systems. \\
\section{ACKNOWLEDGMENT}
The authors acknowledge the work by the cyclotron and ion-source groups at KVI for delivering a high-quality beam used
in these measurements. The present work has been performed with financial support from the ``Nederlandse Organisatie voor Wetenschappelijk Onderzoek'' (NWO). This work was also partly supported by the Polish National Science Centre under Grant No. 2012/05/B/ST2/02556 and 2016/22/M/ST2/00173. A.D. acknowledges support by the Alexander von Humboldt Foundation under Grant No. LTU-1185721-HFST-E. 

%

\begin{figure*}[ht]
\centering
\resizebox{18cm}{!}{\includegraphics[angle = 0,width =1\textwidth]{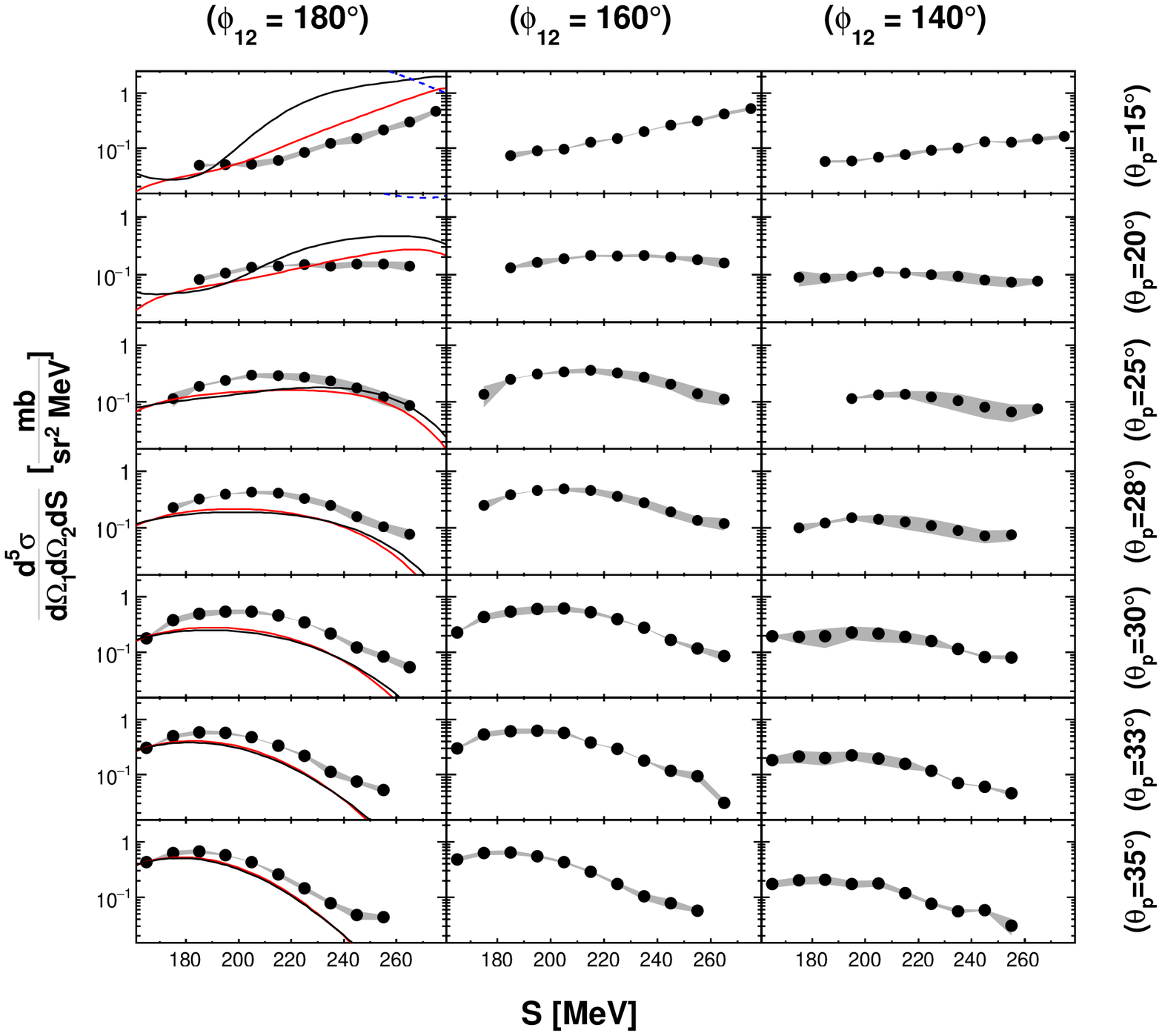}}
\vspace*{-0.4cm}
\caption{Results of differential cross sections as a function of $S$ for the three-body break-up channel of the reaction $^{2}{\rm H}(\vec d,dp)n$ for the configurations for which $\theta_d = 15^\circ$. Other kinematical variables are shown at the top and on the right side of the figure. The gray bands show the systematic uncertainty coming from the calibration procedure, which is dominant in the total systematic uncertainty. The contribution of the other systematic uncertainties are around 7$\%$ of each measured value. Also, the systematic uncertainty due to the luminosity measurement is around 7$\%$ and is not presented here. The red (black) solid line shows the results of the SSA1 (SSA4) approximation using CD-Bonn+$\Delta$ potential and the blue-dashed line indicates the neutron energy using the same scale with a unit of MeV. This line is not visible in all plots.}
\label{Crse15}
\end{figure*}
\begin{figure*}[ht]
\centering
\resizebox{18cm}{!}{\includegraphics[angle = 0,width =1\textwidth]{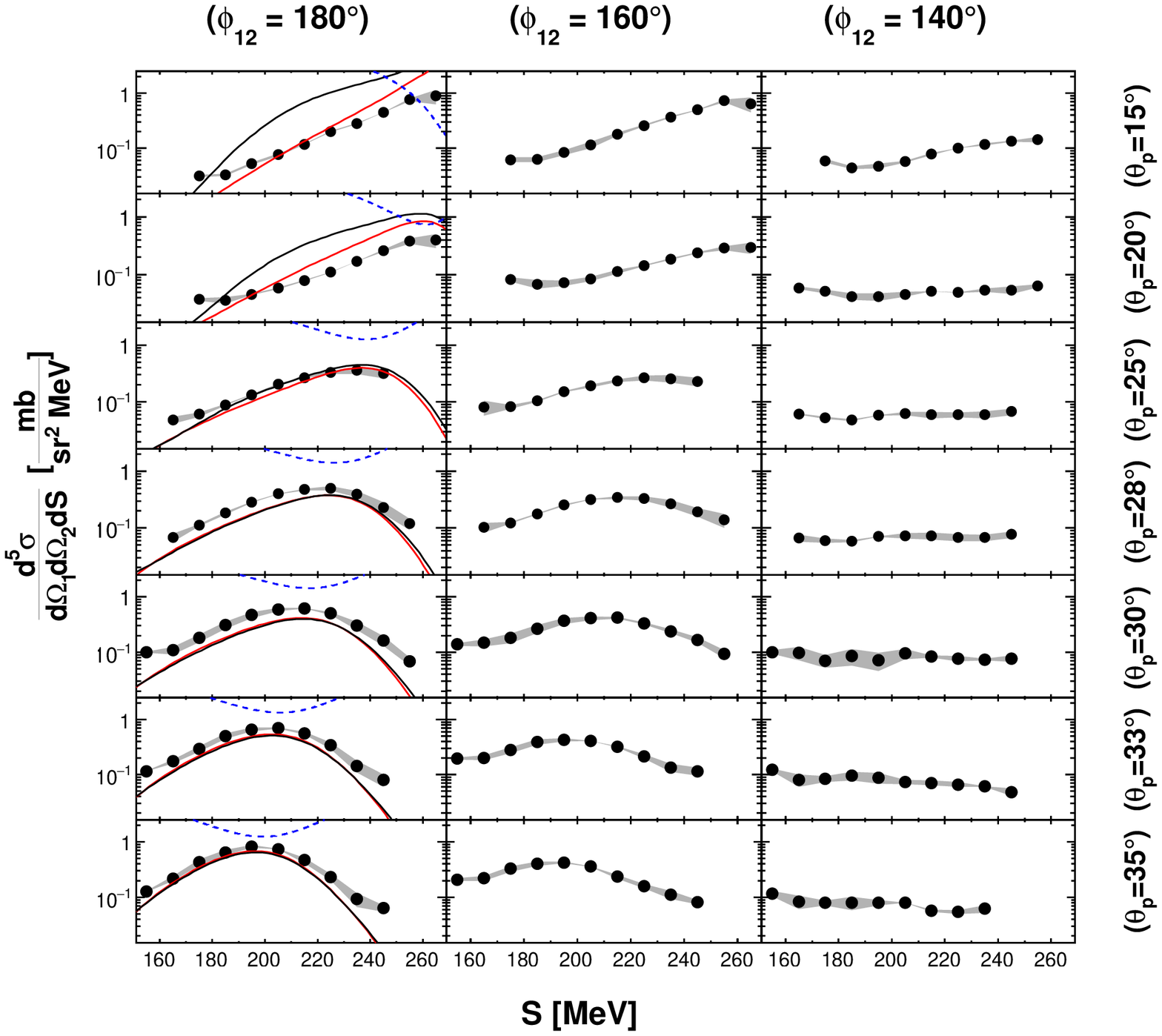}}
\vspace*{-0.4cm}
\caption{Same as Fig.~\ref{Crse15} except for $\theta_d = 20^\circ$.}
\label{Crse20}
\end{figure*}
\begin{figure*}[ht]
\centering
\resizebox{18cm}{!}{\includegraphics[angle = 0,width =1\textwidth]{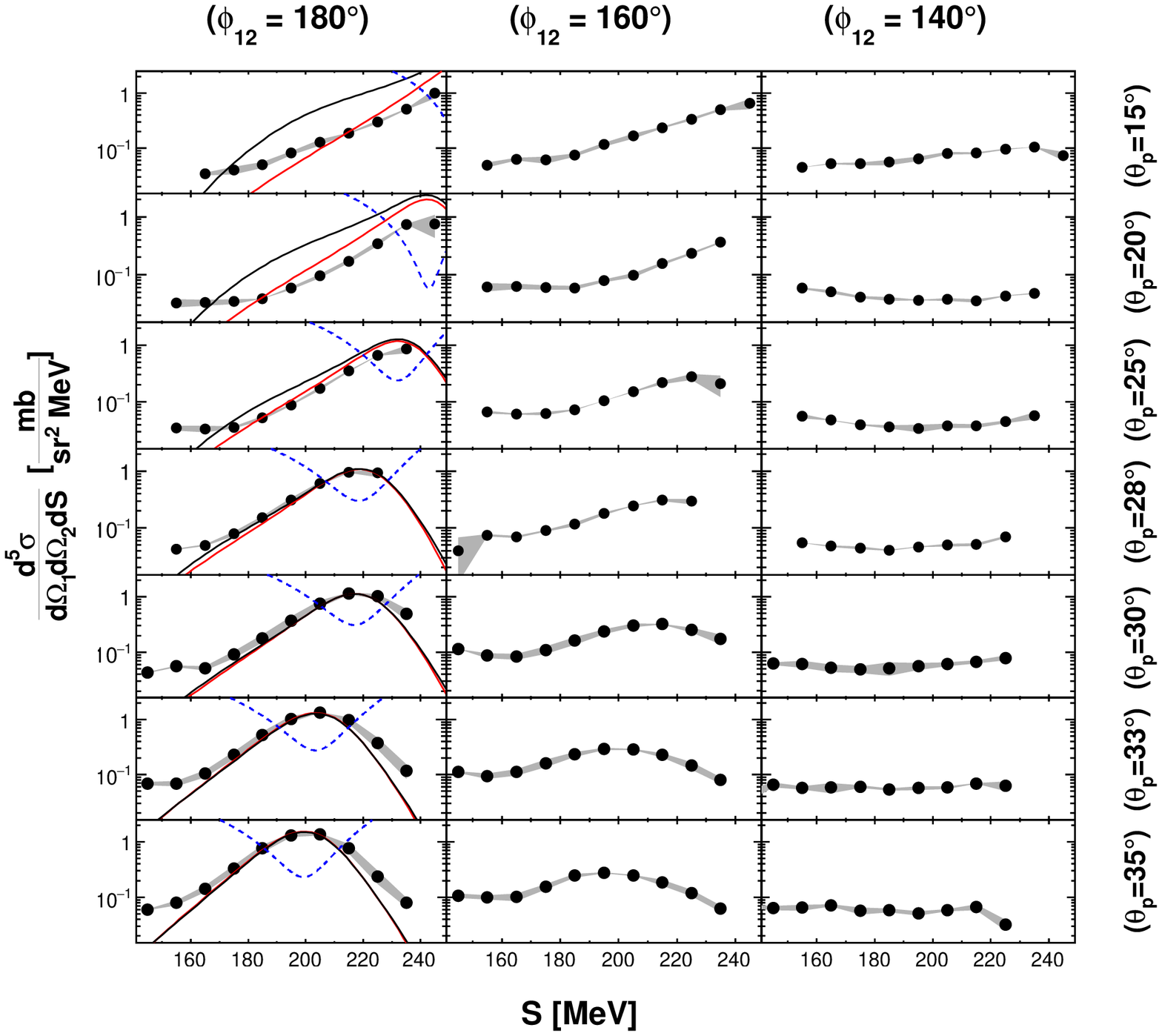}}
\vspace*{-0.4cm}
\caption{Same as Fig.~\ref{Crse15} except for $\theta_d = 25^\circ$.}
\label{Crse25}
\end{figure*}
\begin{figure*}[ht]
\centering
\resizebox{18cm}{!}{\includegraphics[angle = 0,width =1\textwidth]{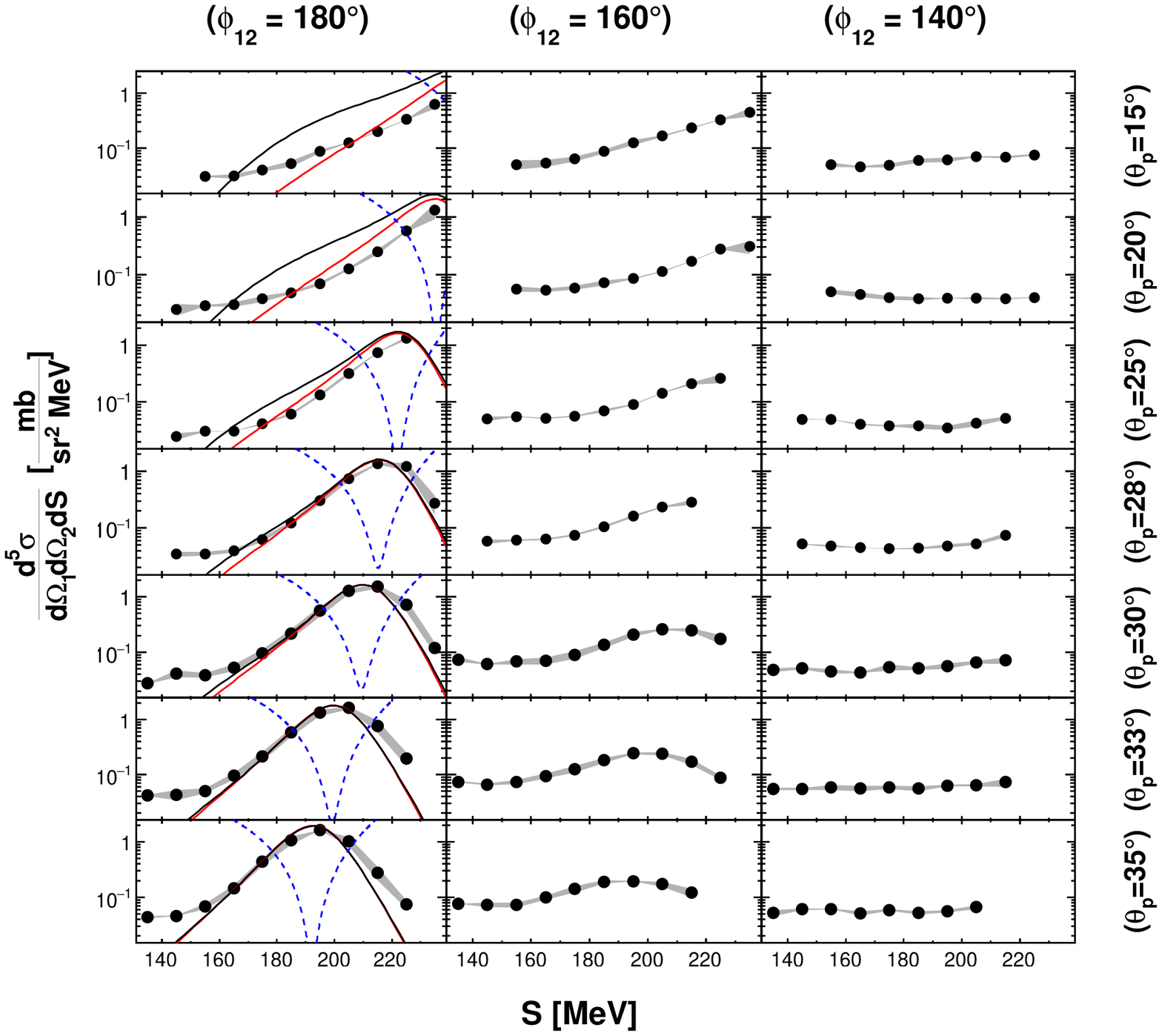}}
\vspace*{-0.4cm}
\caption{Same as Fig.~\ref{Crse15} except for $\theta_d = 28^\circ$.}
\label{Crse28}
\end{figure*}

\begin{figure*}[ht]
\centering
\resizebox{18cm}{!}{\includegraphics[angle = 0,width =1\textwidth]{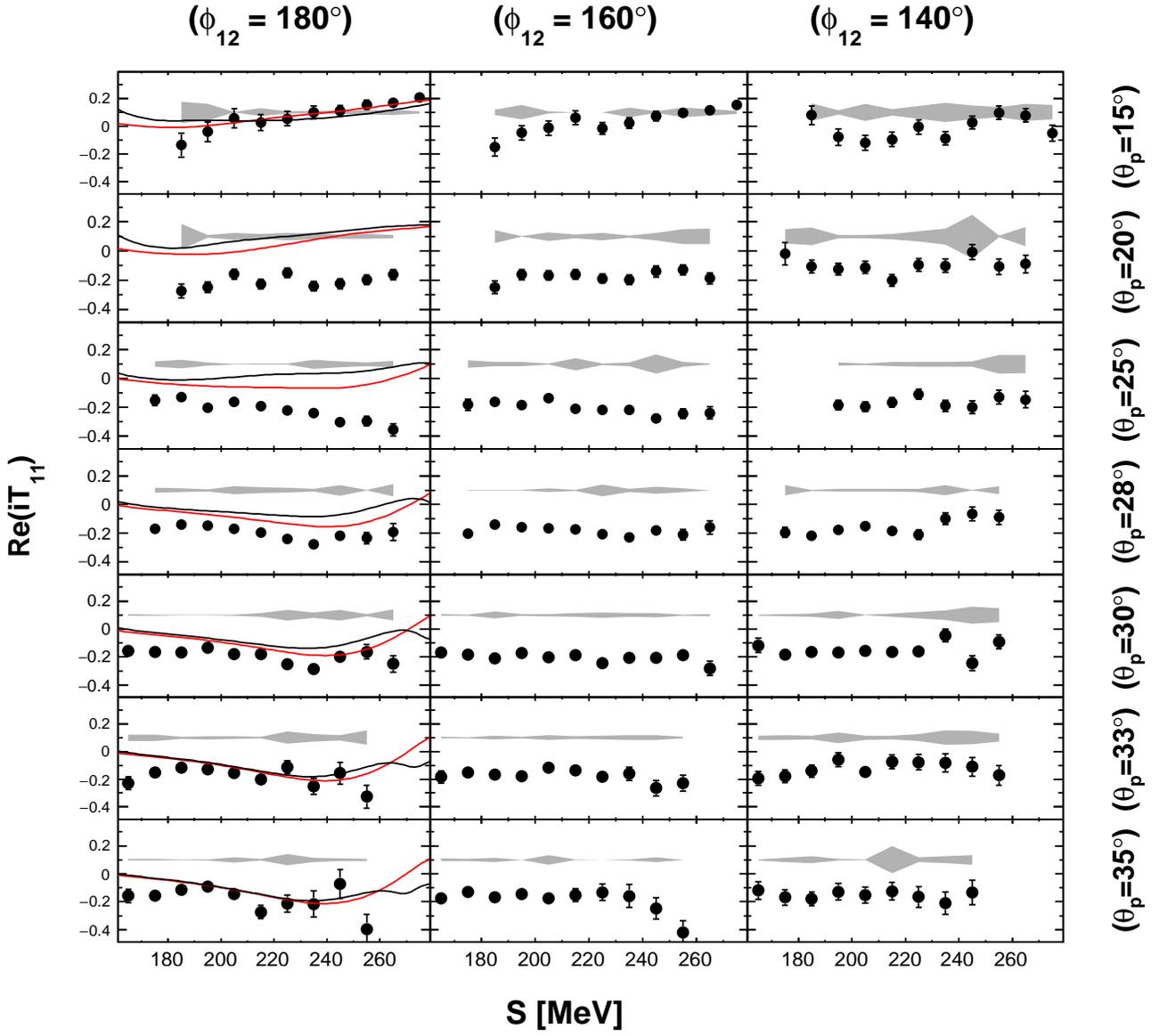}}
\vspace*{-0.4cm}
\caption{The results of $Re(iT_{11})$ with the same information as in Fig.~\ref{Crse15} except that the contribution of the other systematic uncertainties are around 5$\%$ of each measured value.}
\label{iT11_15}
\end{figure*}

\begin{figure*}[ht]
\centering
\resizebox{18cm}{!}{\includegraphics[angle = 0,width =1\textwidth]{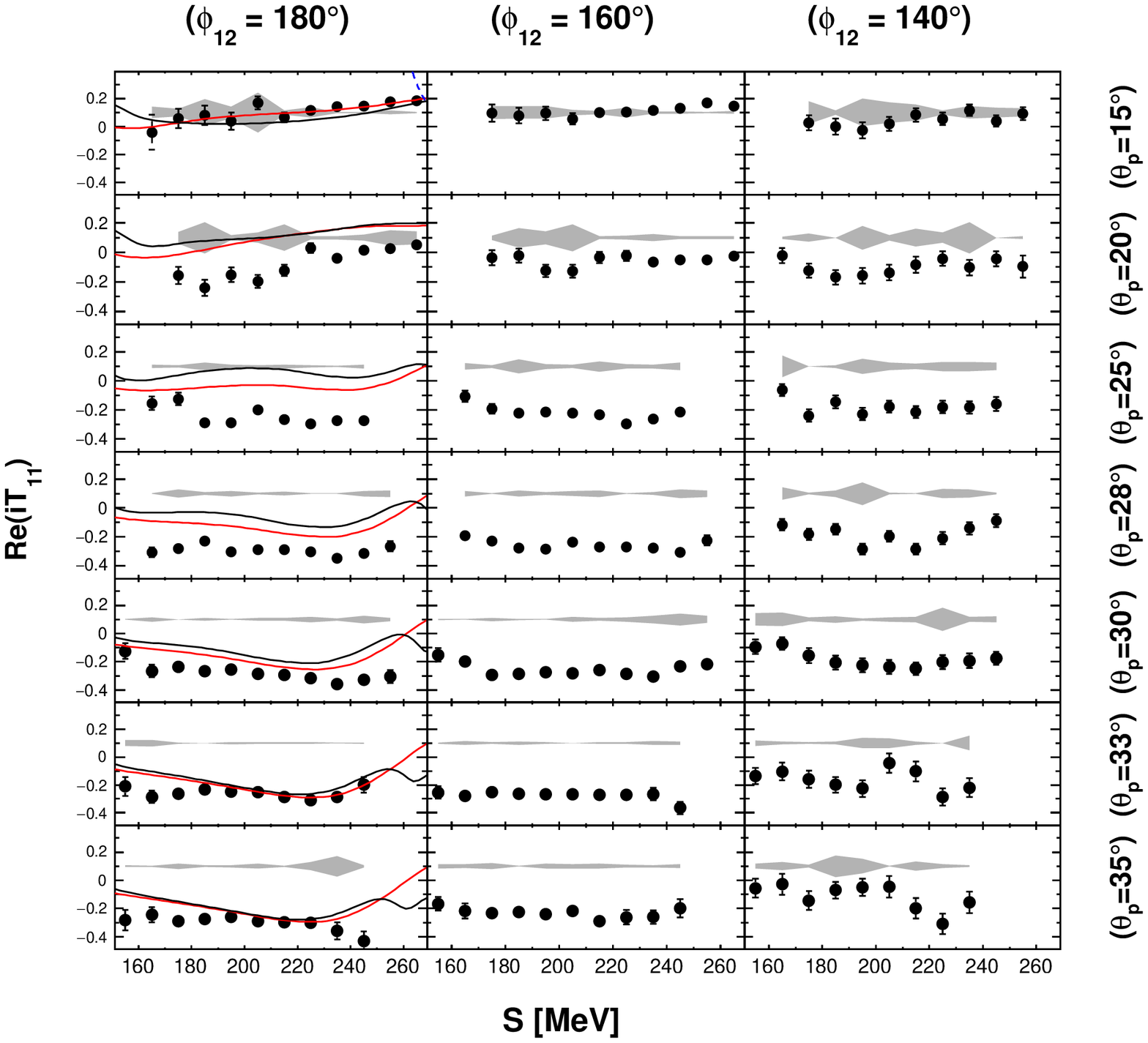}}
\vspace*{-0.4cm}
\caption{Same as Fig.~\ref{iT11_15} except for $\theta_d = 20^\circ$.}
\label{iT11_20}
\end{figure*}
\begin{figure*}[ht]
\centering
\resizebox{18cm}{!}{\includegraphics[angle = 0,width =1\textwidth]{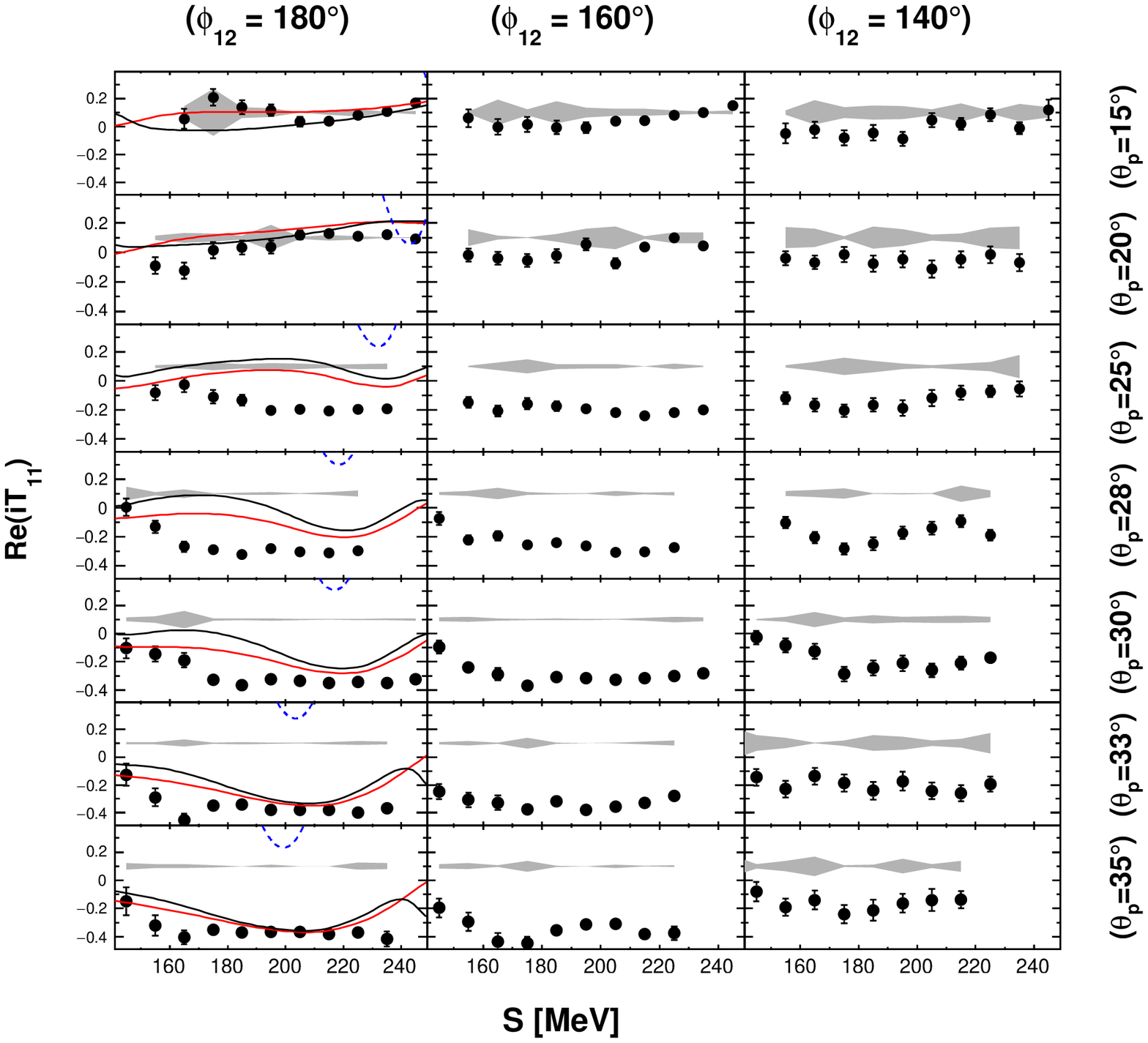}}
\vspace*{-0.4cm}
\caption{Same as Fig.~\ref{iT11_15} except for $\theta_d = 25^\circ$.}
\label{iT11_25}
\end{figure*}
\begin{figure*}[ht]
\centering
\resizebox{18cm}{!}{\includegraphics[angle = 0,width =1\textwidth]{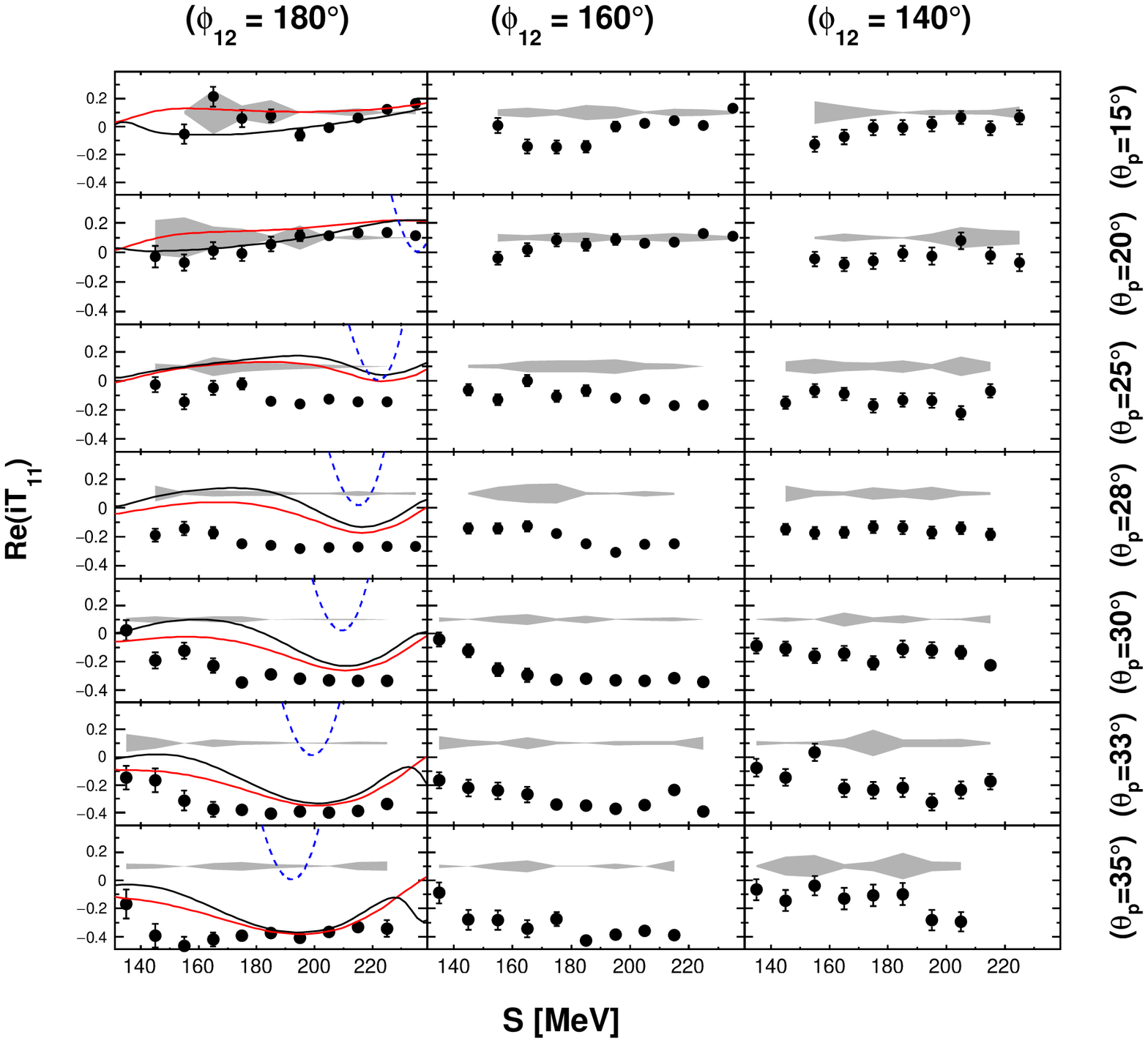}}
\vspace*{-0.4cm}
\caption{Same as Fig.~\ref{iT11_15} except for $\theta_d = 28^\circ$.}
\label{iT11_28}
\end{figure*}
\begin{figure*}[ht]
\centering
\resizebox{18cm}{!}{\includegraphics[angle = 0,width =1\textwidth]{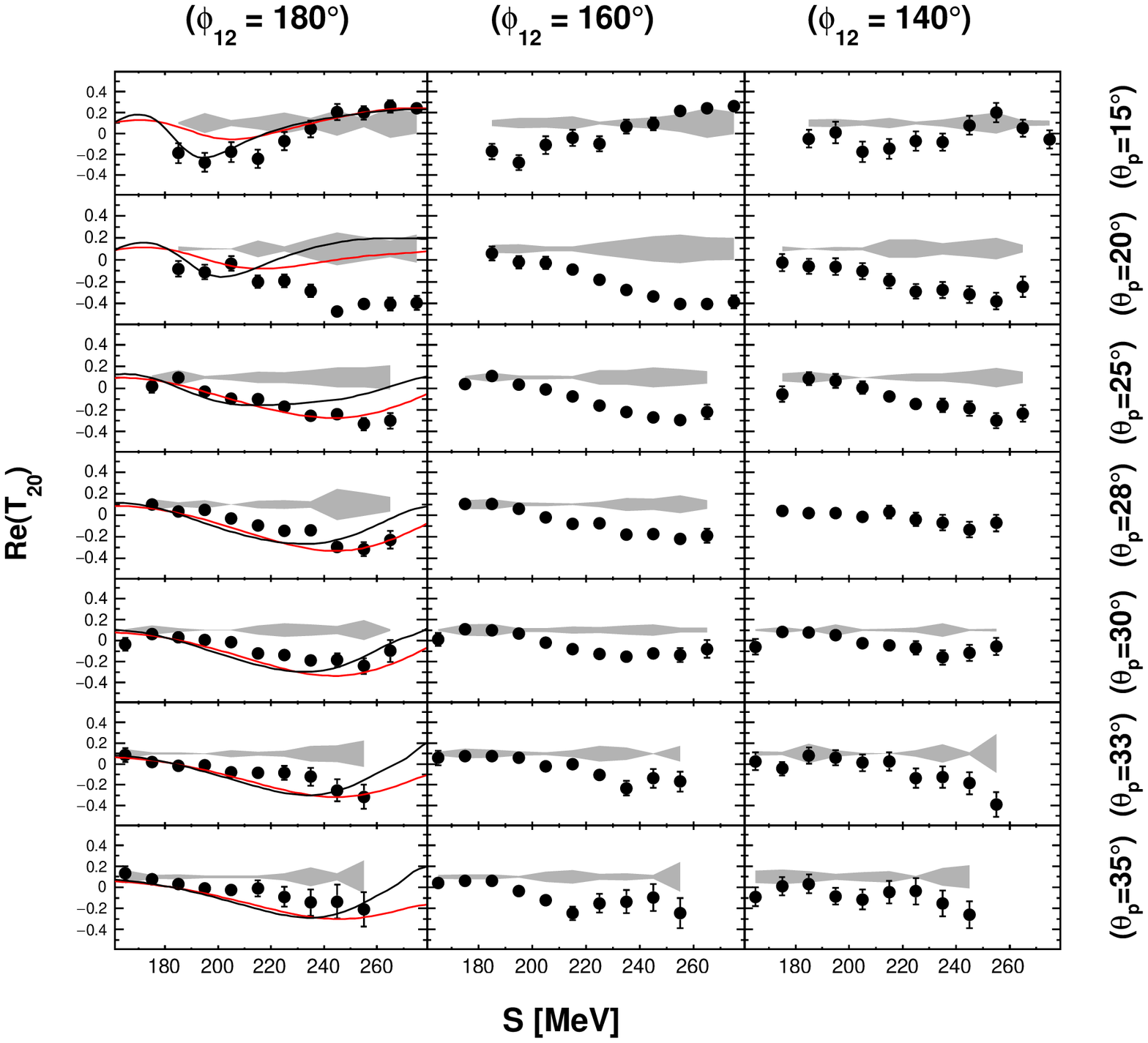}}
\vspace*{-0.4cm}
\caption{The results of $Re(T_{20})$ with the same information as in Fig.~\ref{iT11_15}.}
\label{T20_15}
\end{figure*}
\begin{figure*}[ht]
\centering
\resizebox{18cm}{!}{\includegraphics[angle = 0,width =1\textwidth]{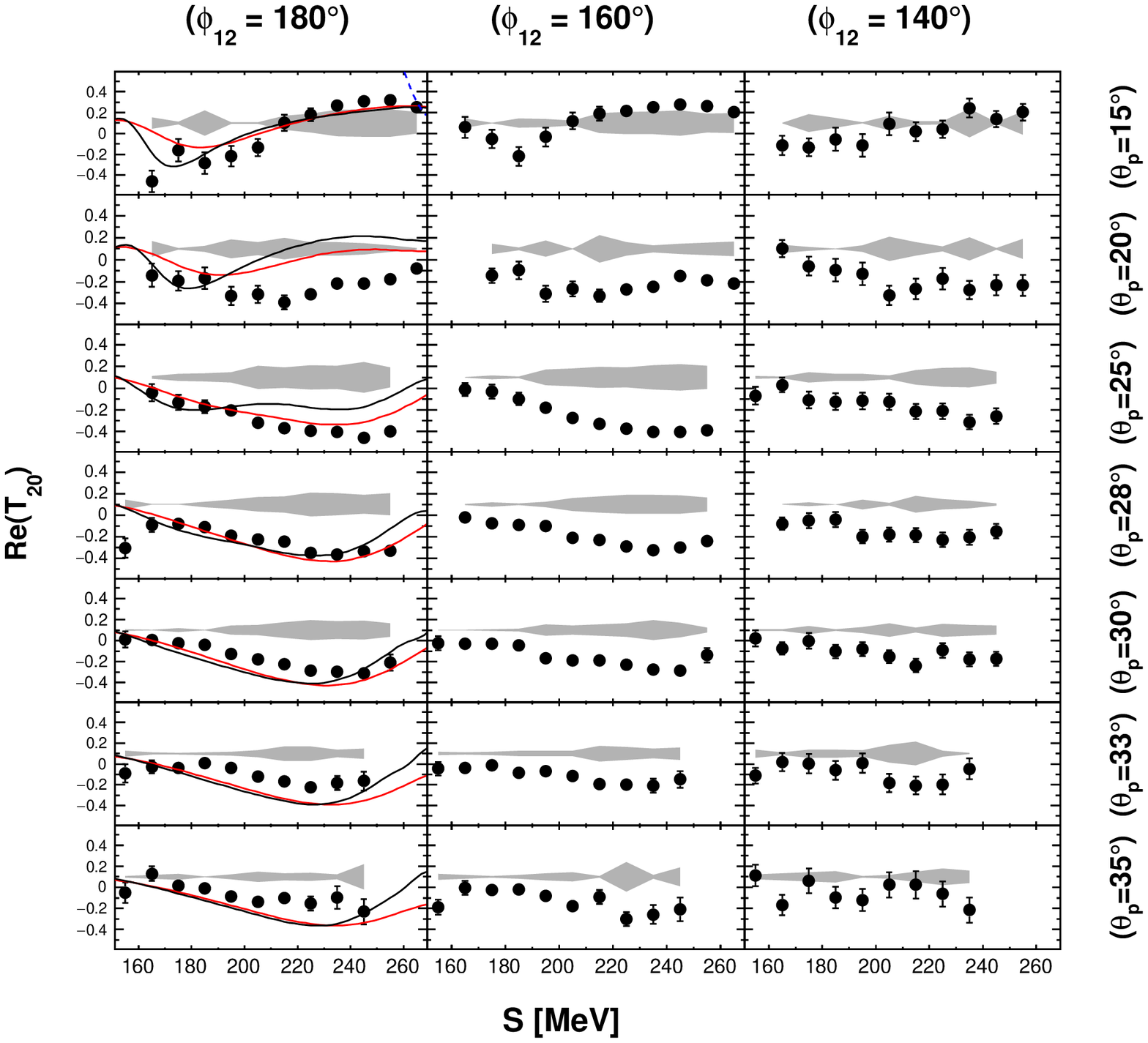}}
\vspace*{-0.4cm}
\caption{Same as Fig.~\ref{T20_15} except for $\theta_d = 20^\circ$.}
\label{T20_20}
\end{figure*}
\begin{figure*}[ht]
\centering
\resizebox{18cm}{!}{\includegraphics[angle = 0,width =1\textwidth]{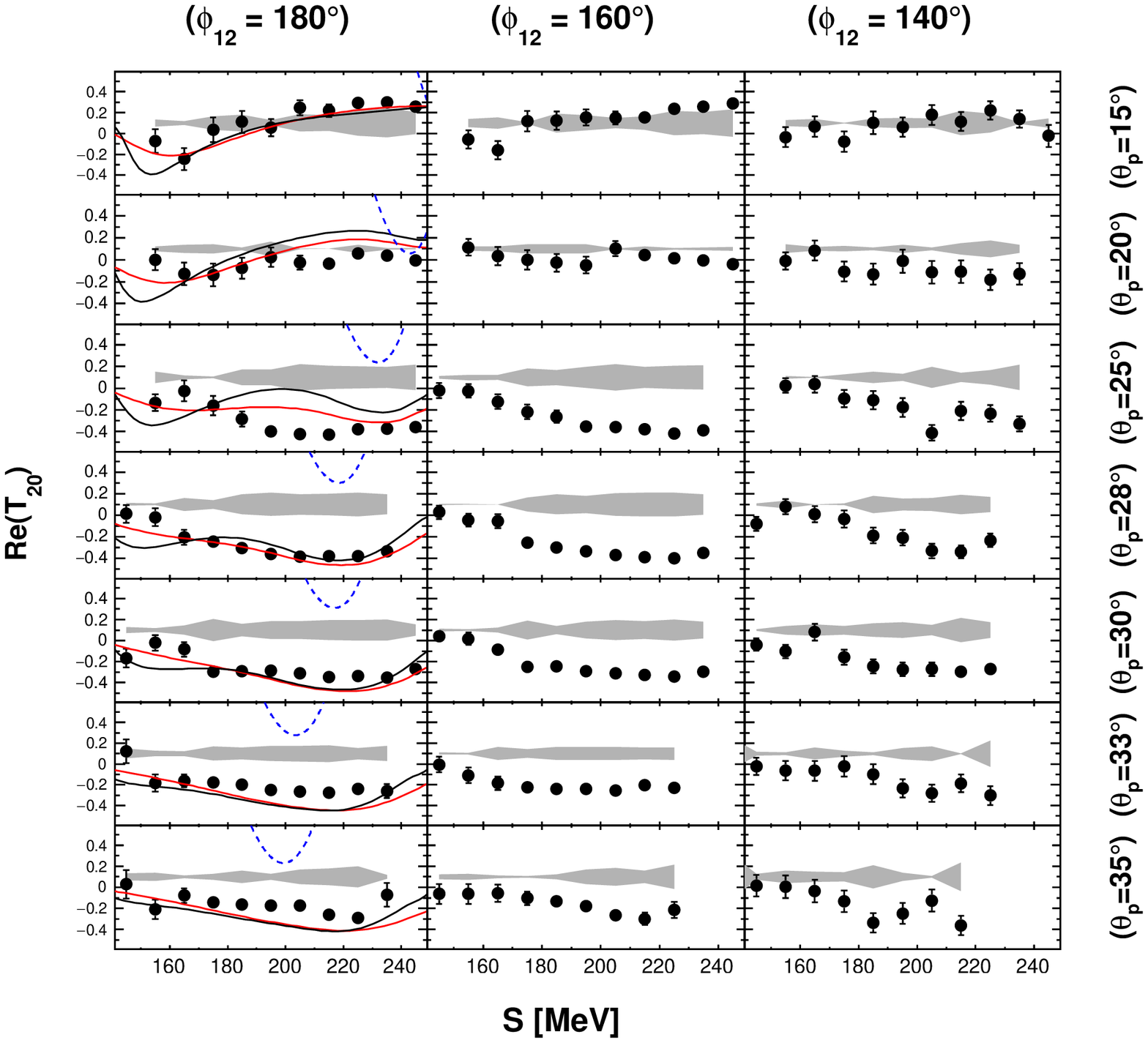}}
\vspace*{-0.4cm}
\caption{Same as Fig.~\ref{T20_15} except for $\theta_d = 25^\circ$.}
\label{T20_25}
\end{figure*}
\begin{figure*}[ht]
\centering
\resizebox{18cm}{!}{\includegraphics[angle = 0,width =1\textwidth]{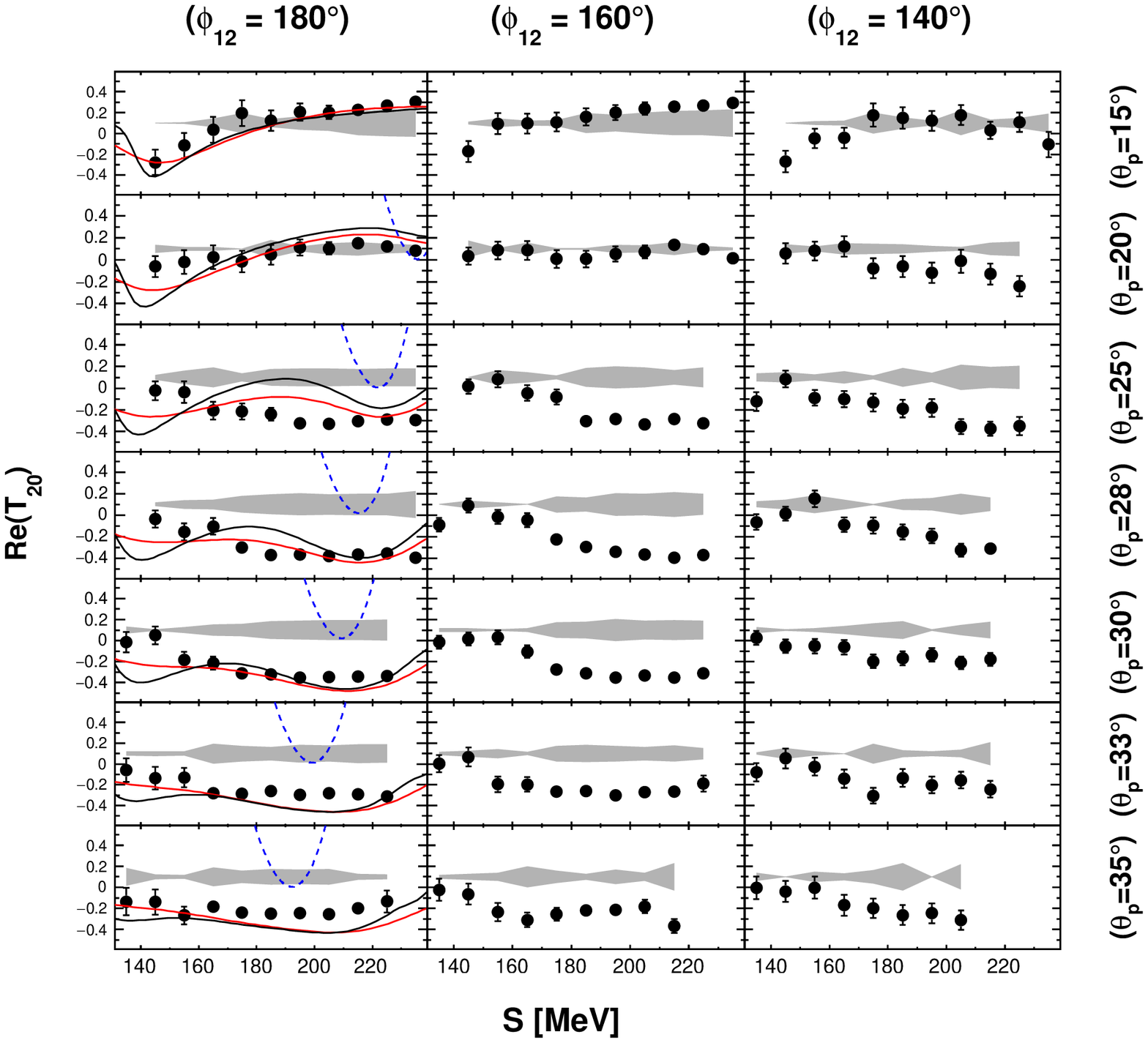}}
\vspace*{-0.4cm}
\caption{Same as Fig.~\ref{T20_15} except for $\theta_d = 28^\circ$.}
\label{T20_28}
\end{figure*}
\begin{figure*}[ht]
\centering
\resizebox{18cm}{!}{\includegraphics[angle = 0,width =1\textwidth]{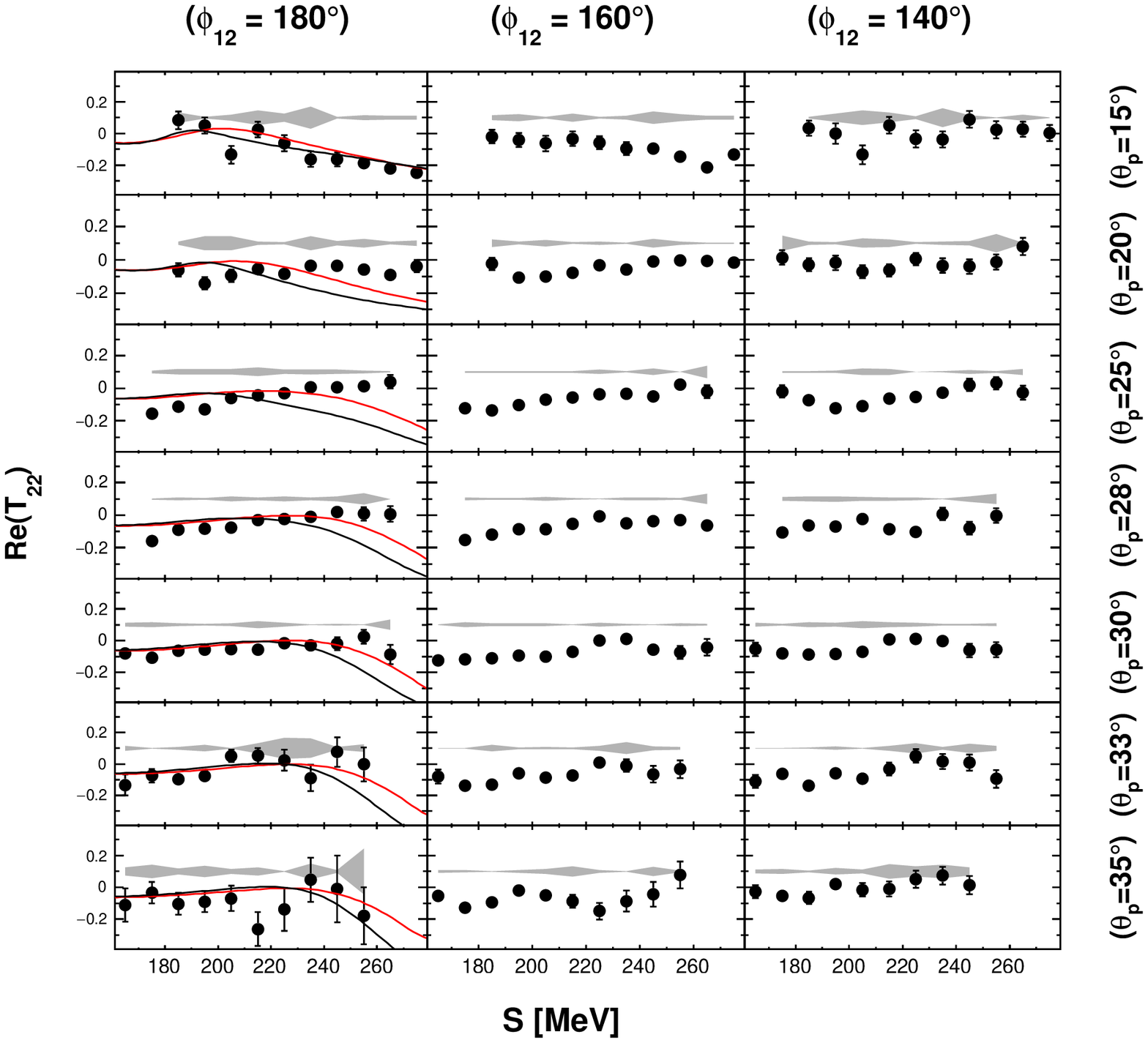}}
\vspace*{-0.4cm}
\caption{The results of $Re(T_{22})$ with the same information as in Fig.~\ref{iT11_15}.}
\label{T22_15}
\end{figure*}
\begin{figure*}[ht]
\centering
\resizebox{18cm}{!}{\includegraphics[angle = 0,width =1\textwidth]{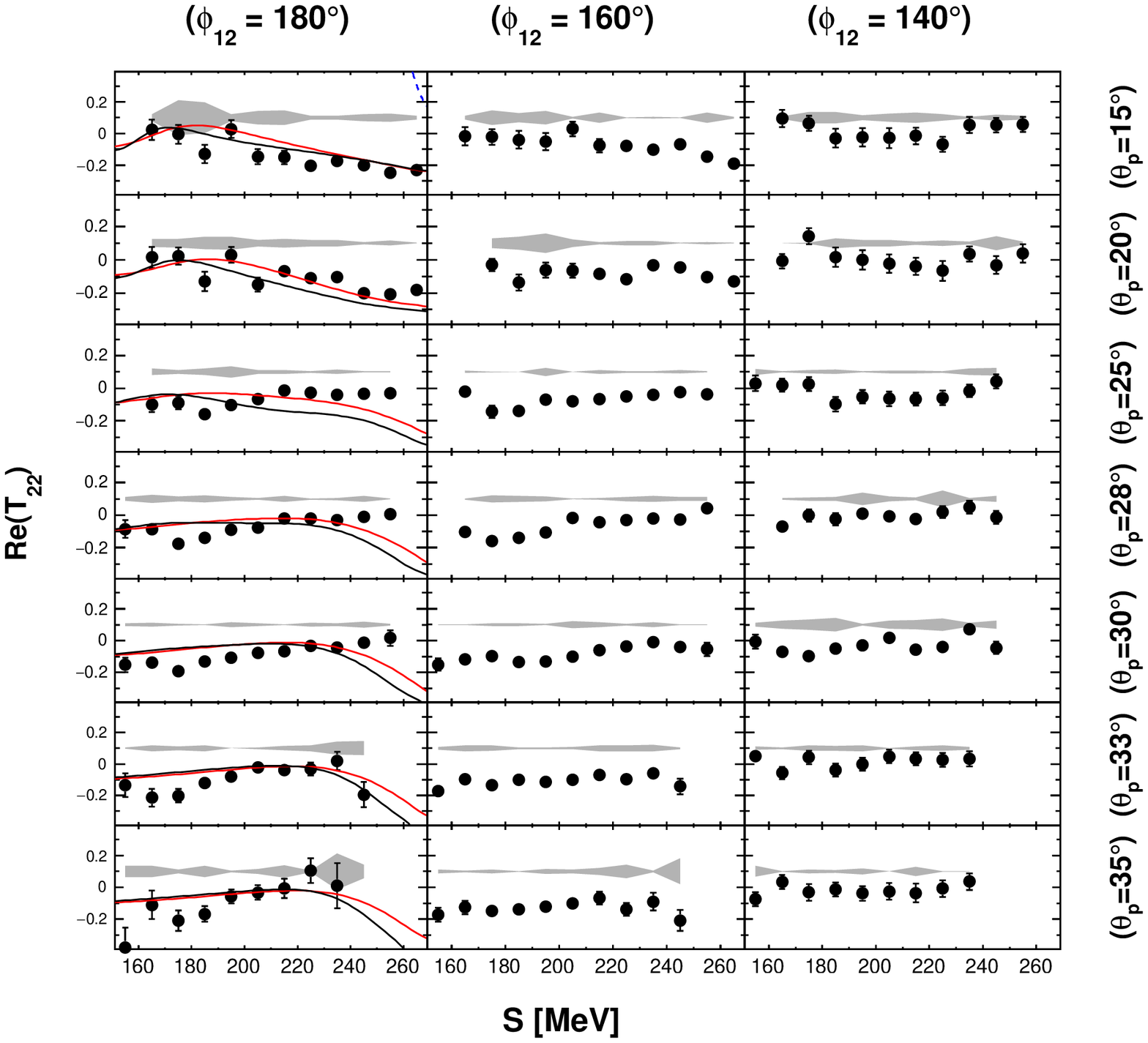}}
\vspace*{-0.4cm}
\caption{Same as Fig.~\ref{T22_15} except for $\theta_d = 20^\circ$.}
\label{T22_20}
\end{figure*}
\begin{figure*}[ht]
\centering
\resizebox{18cm}{!}{\includegraphics[angle = 0,width =1\textwidth]{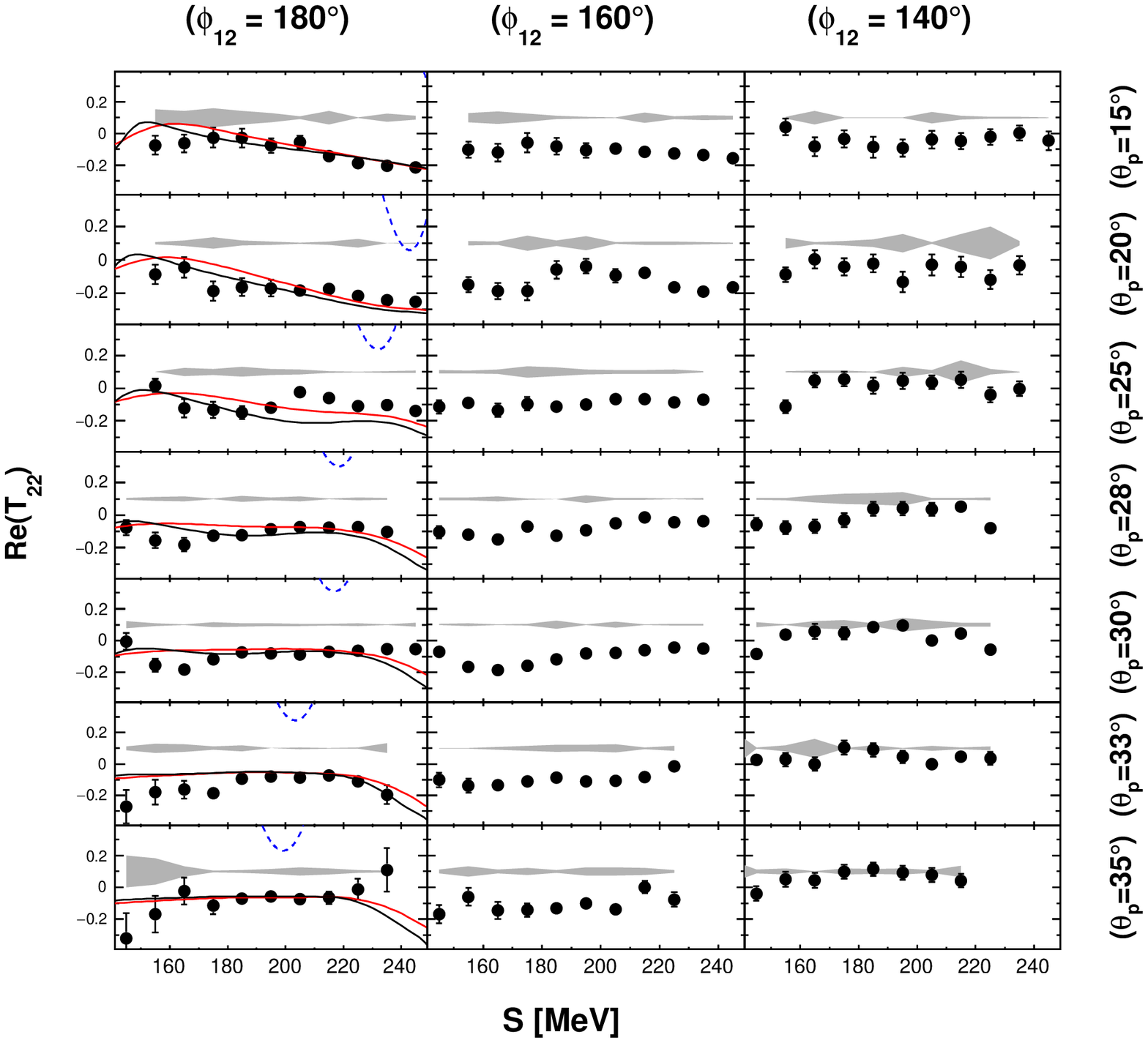}}
\vspace*{-0.4cm}
\caption{Same as Fig.~\ref{T22_15} except for $\theta_d = 25^\circ$.}
\label{T22_25}
\end{figure*}
\begin{figure*}[ht]
\centering
\resizebox{18cm}{!}{\includegraphics[angle = 0,width =1\textwidth]{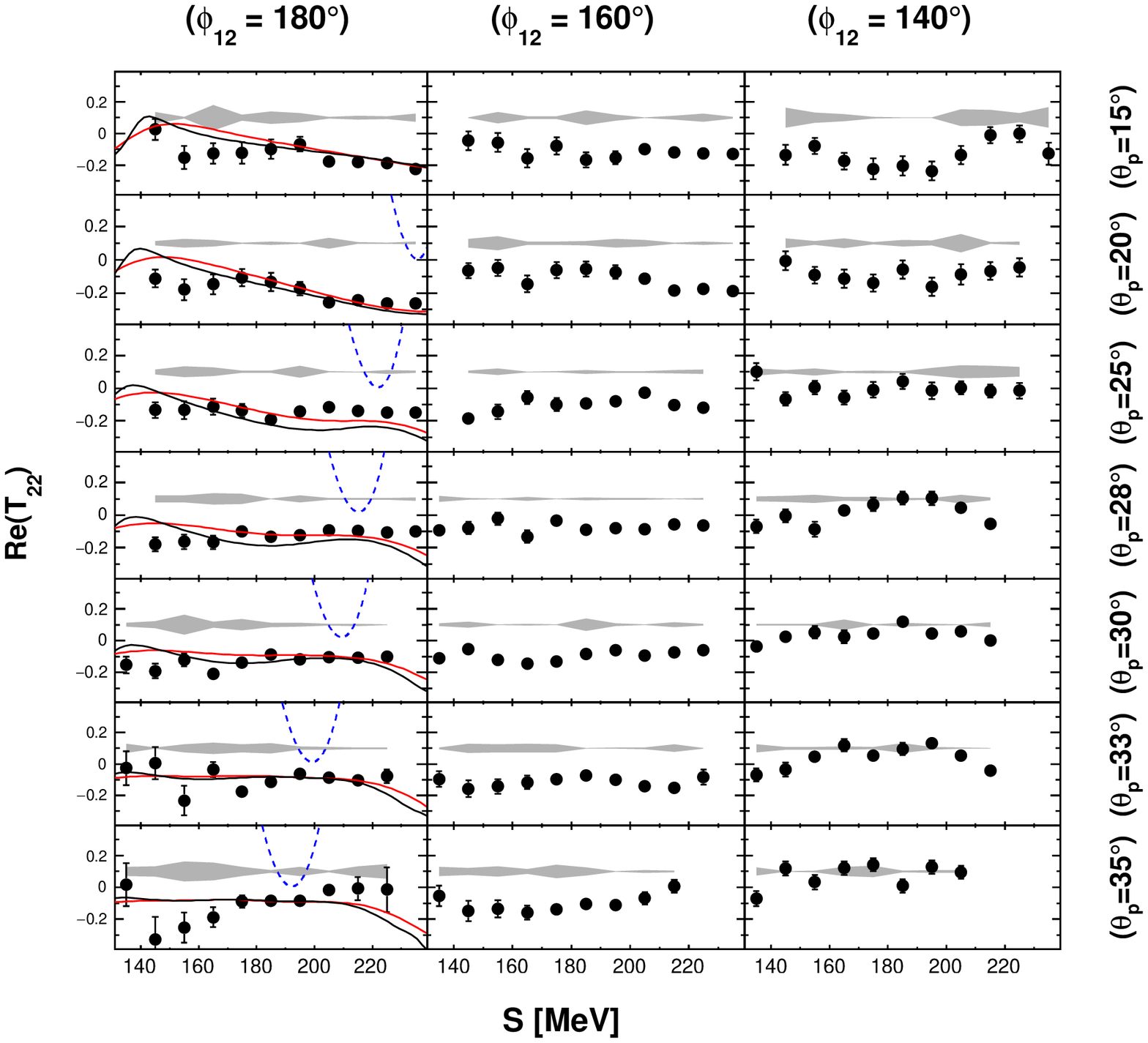}}
\vspace*{-0.4cm}
\caption{Same as Fig.~\ref{T22_15} except for $\theta_d = 28^\circ$.}
\label{T22_28}
\end{figure*}
\clearpage
\begin{figure*}[ht]
\centering
\resizebox{18cm}{!}{\includegraphics[angle = 0,width =1\textwidth]{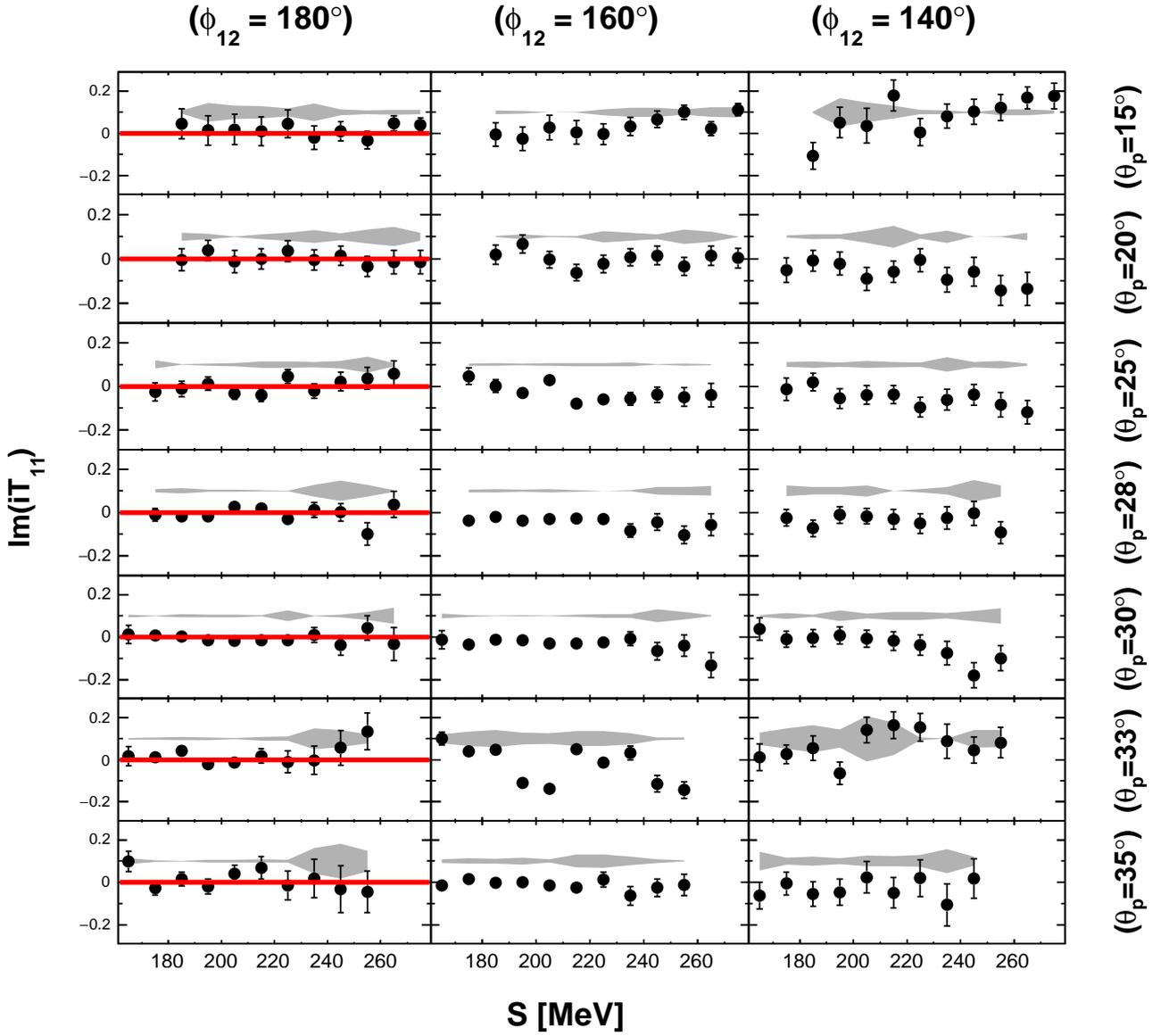}}
\vspace*{-0.4cm}
\caption{The results of $Im(iT_{11})$ as a function of $S$ for the three-body break-up channel of the $^{2}{\rm H}(\vec d,dp)n$ reaction for the configurations for which $\theta_d = 15^\circ$. Other kinematical variables are shown at the top and on the right side of the figure. The solid lines represent the zero line. The gray bands show the systematic uncertainty coming from the calibration procedure.}
\label{ImiT11_15}
\end{figure*}
\begin{figure*}[ht]
\centering
\resizebox{18cm}{!}{\includegraphics[angle = 0,width =1\textwidth]{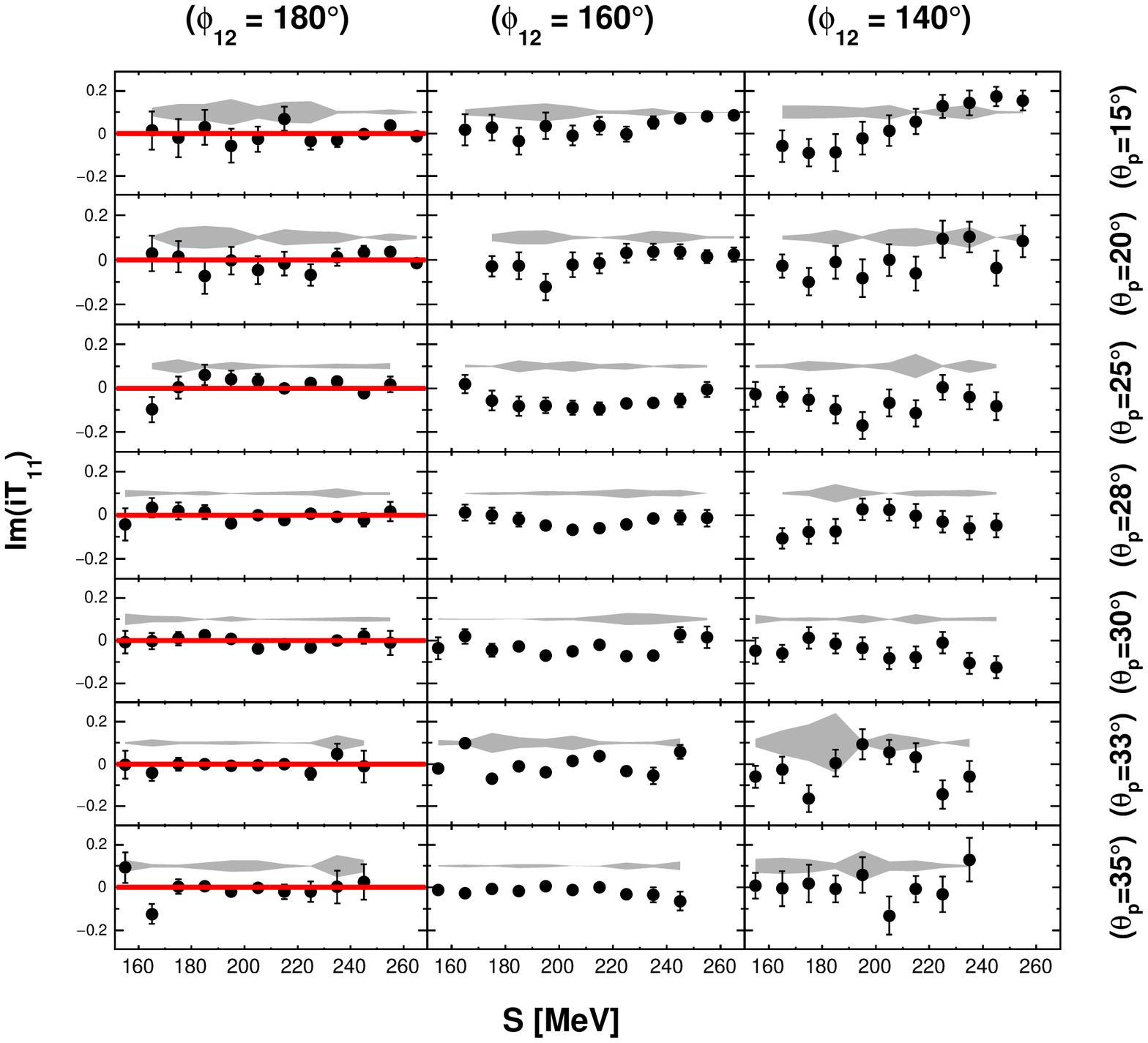}}
\vspace*{-0.4cm}
\caption{Same as Fig.~\ref{ImiT11_15} except for $\theta_d = 20^\circ$.}
\label{ImiT11_20}
\end{figure*}

\begin{figure*}[ht]
\centering
\resizebox{18cm}{!}{\includegraphics[angle = 0,width =1\textwidth]{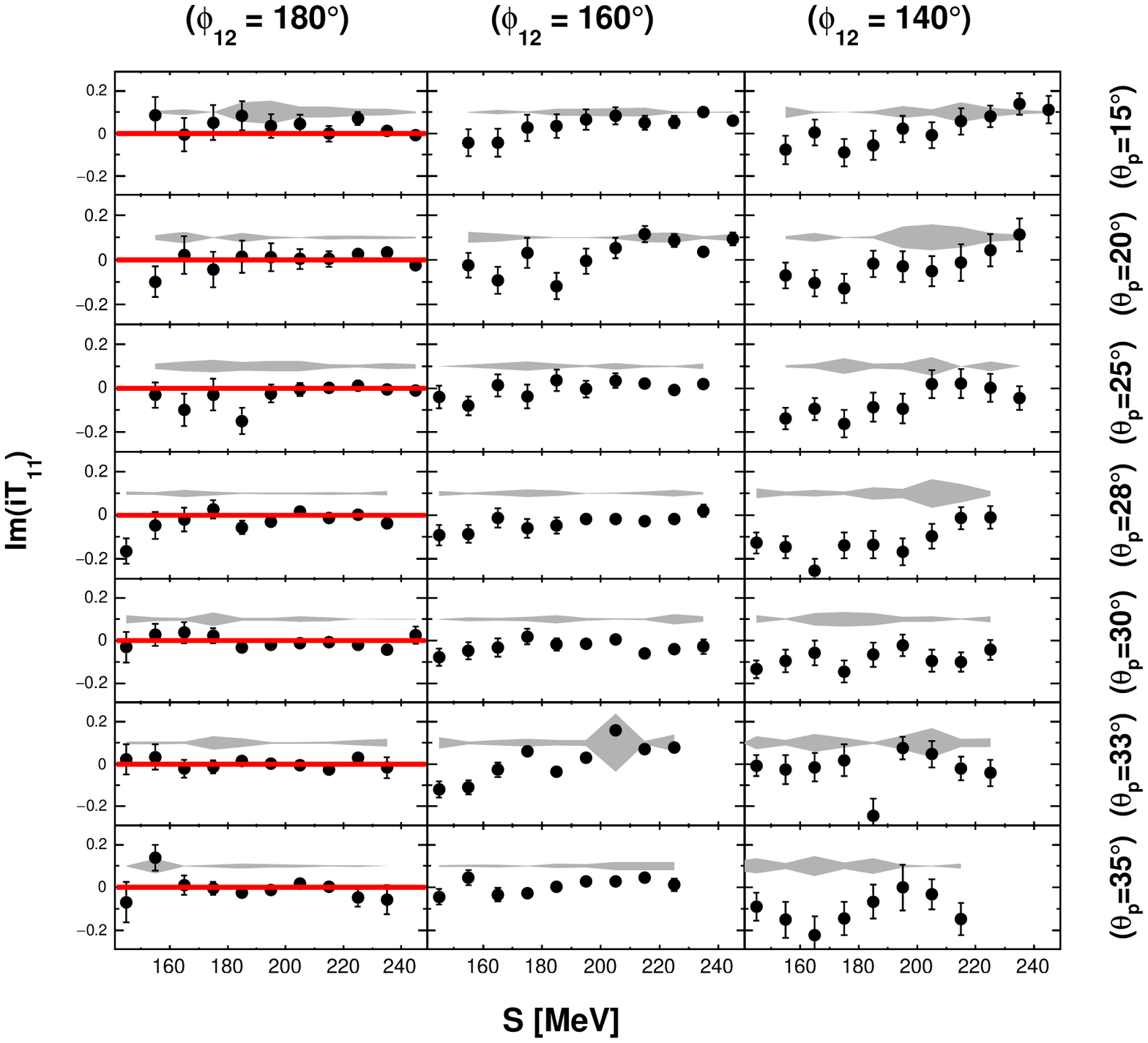}}
\vspace*{-0.4cm}
\caption{Same as Fig.~\ref{ImiT11_15} except for $\theta_d = 25^\circ$.}
\label{ImiT11_25}
\end{figure*}
\begin{figure*}[ht]
\centering
\resizebox{18cm}{!}{\includegraphics[angle = 0,width =1\textwidth]{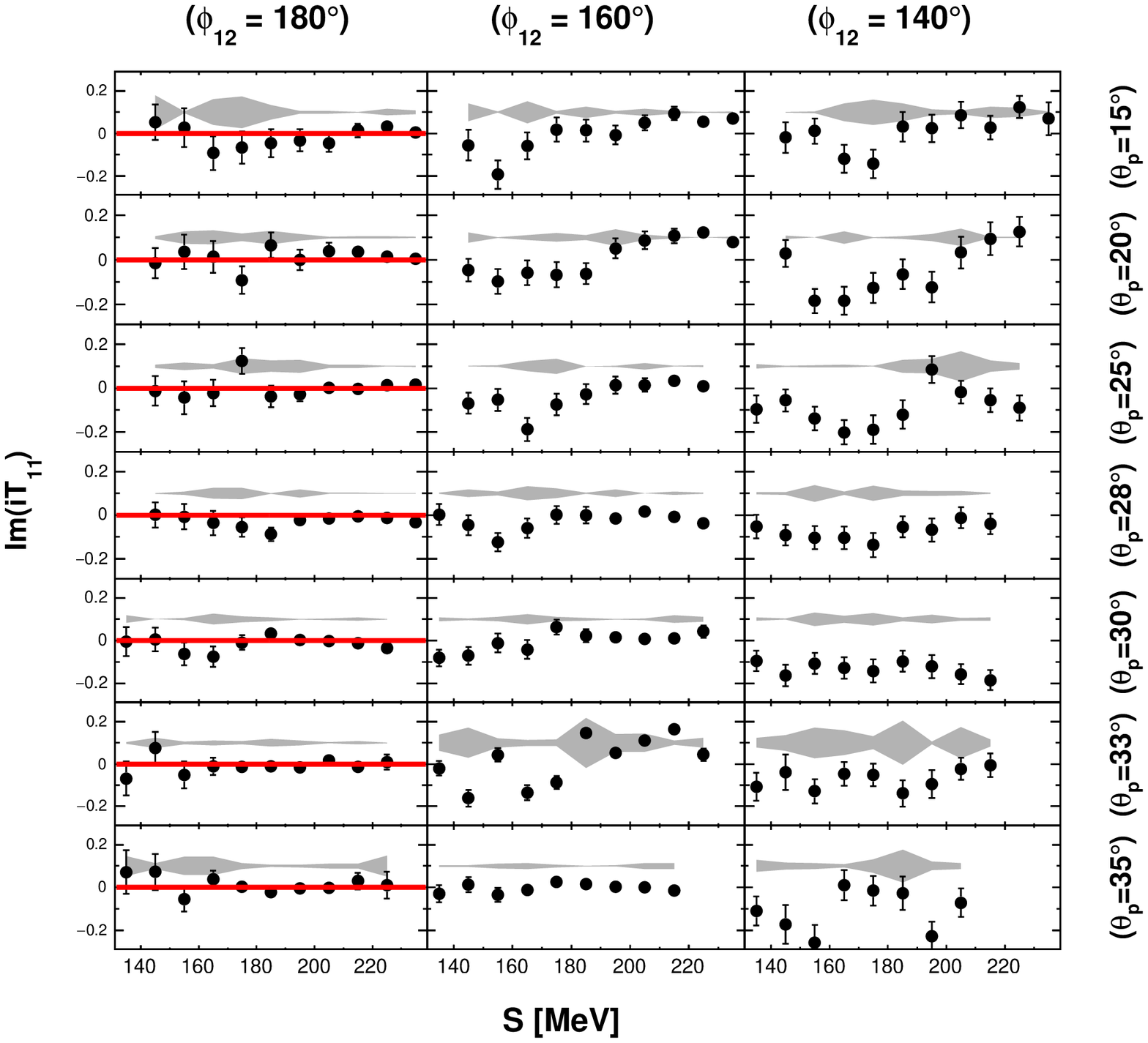}}
\vspace*{-0.4cm}
\caption{Same as Fig.~\ref{ImiT11_15} except for $\theta_d = 28^\circ$.}
\label{ImiT11_28}
\end{figure*}

\begin{figure*}[ht]
\centering
\resizebox{18cm}{!}{\includegraphics[angle = 0,width =1\textwidth]{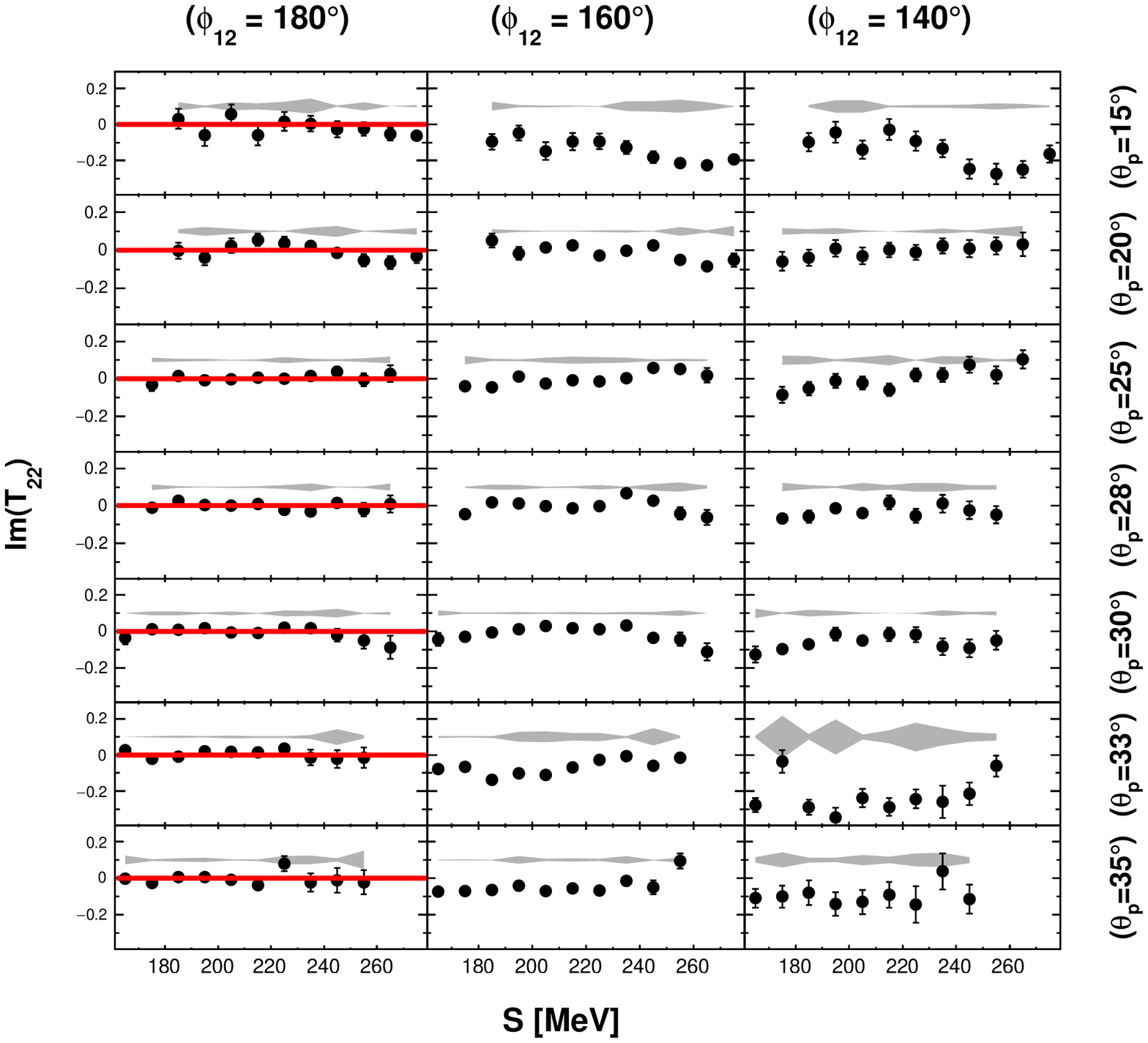}}
\vspace*{-0.4cm}
\caption{The results of $Im(T_{22})$ with the same information as in Fig.~\ref{ImiT11_15}.}
\label{ImT22_15}
\end{figure*}
\begin{figure*}[ht]
\centering
\resizebox{18cm}{!}{\includegraphics[angle = 0,width =1\textwidth]{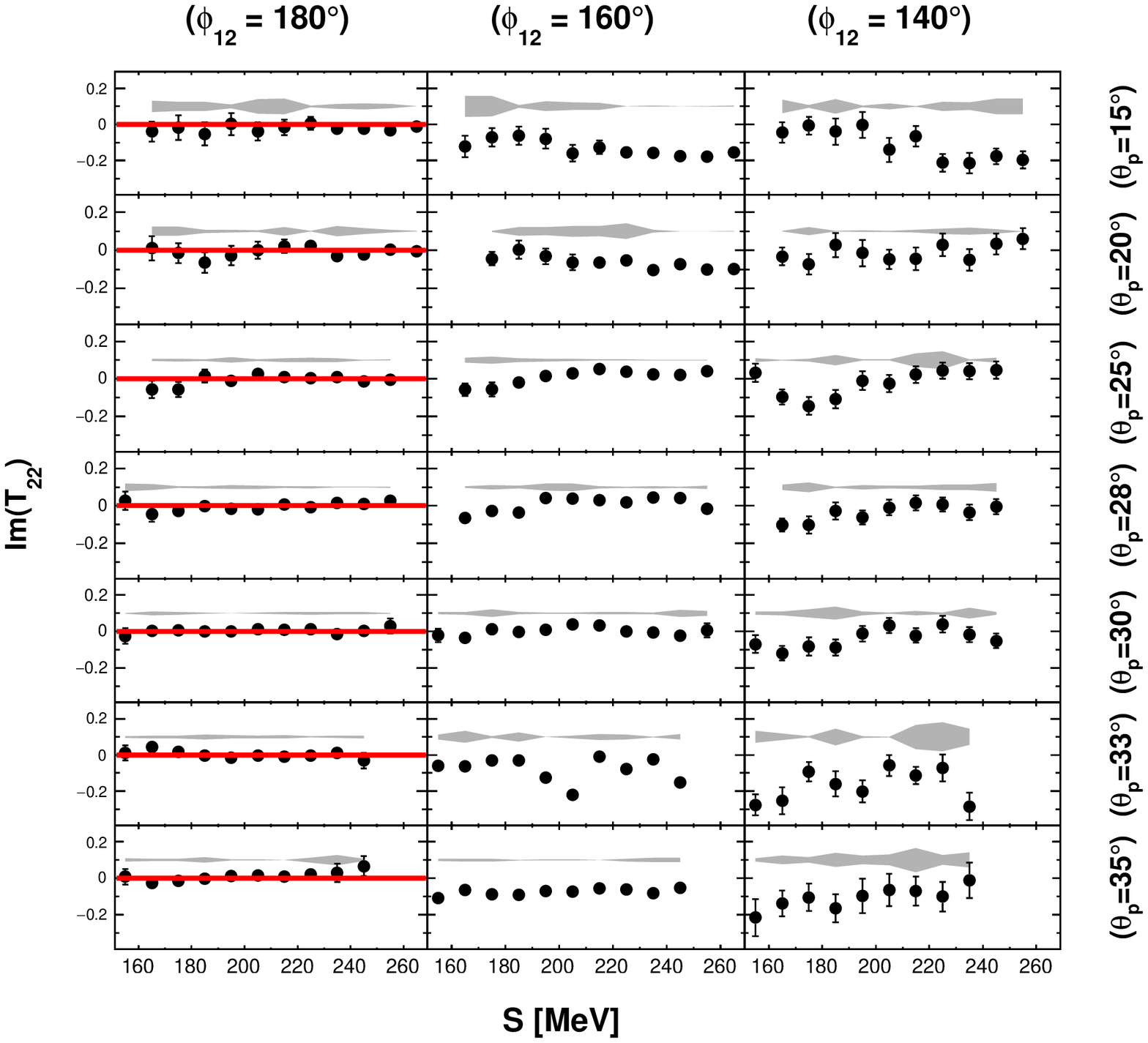}}
\vspace*{-0.4cm}
\caption{Same as Fig.~\ref{ImT22_15} except for $\theta_d = 20^\circ$.}
\label{ImT22_20}
\end{figure*}

\begin{figure*}[ht]
\centering
\resizebox{18cm}{!}{\includegraphics[angle = 0,width =1\textwidth]{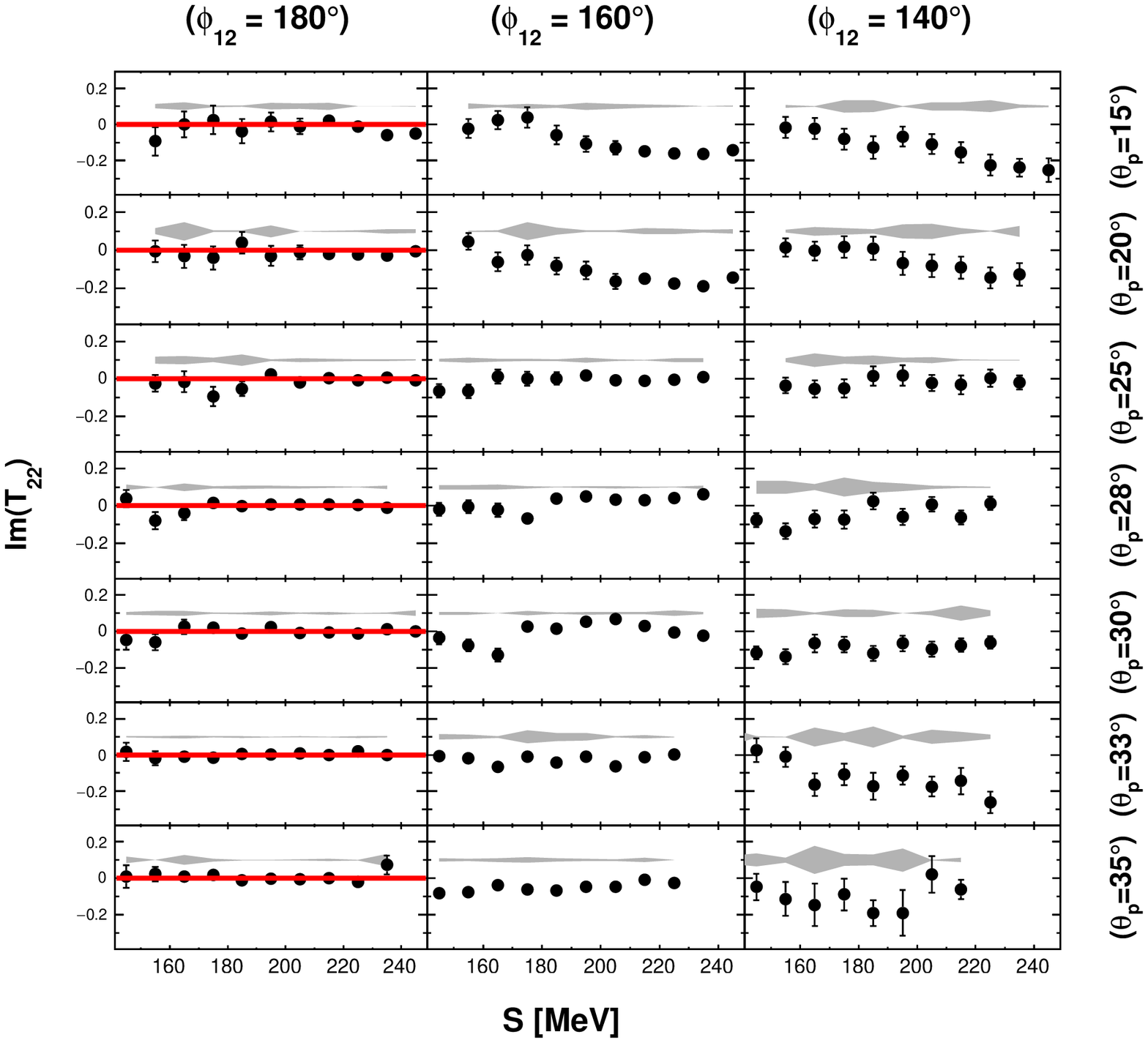}}
\vspace*{-0.4cm}
\caption{Same as Fig.~\ref{ImT22_15} except for $\theta_d = 25^\circ$.}
\label{ImT22_25}
\end{figure*}
\begin{figure*}[ht]
\centering
\resizebox{18cm}{!}{\includegraphics[angle = 0,width =1\textwidth]{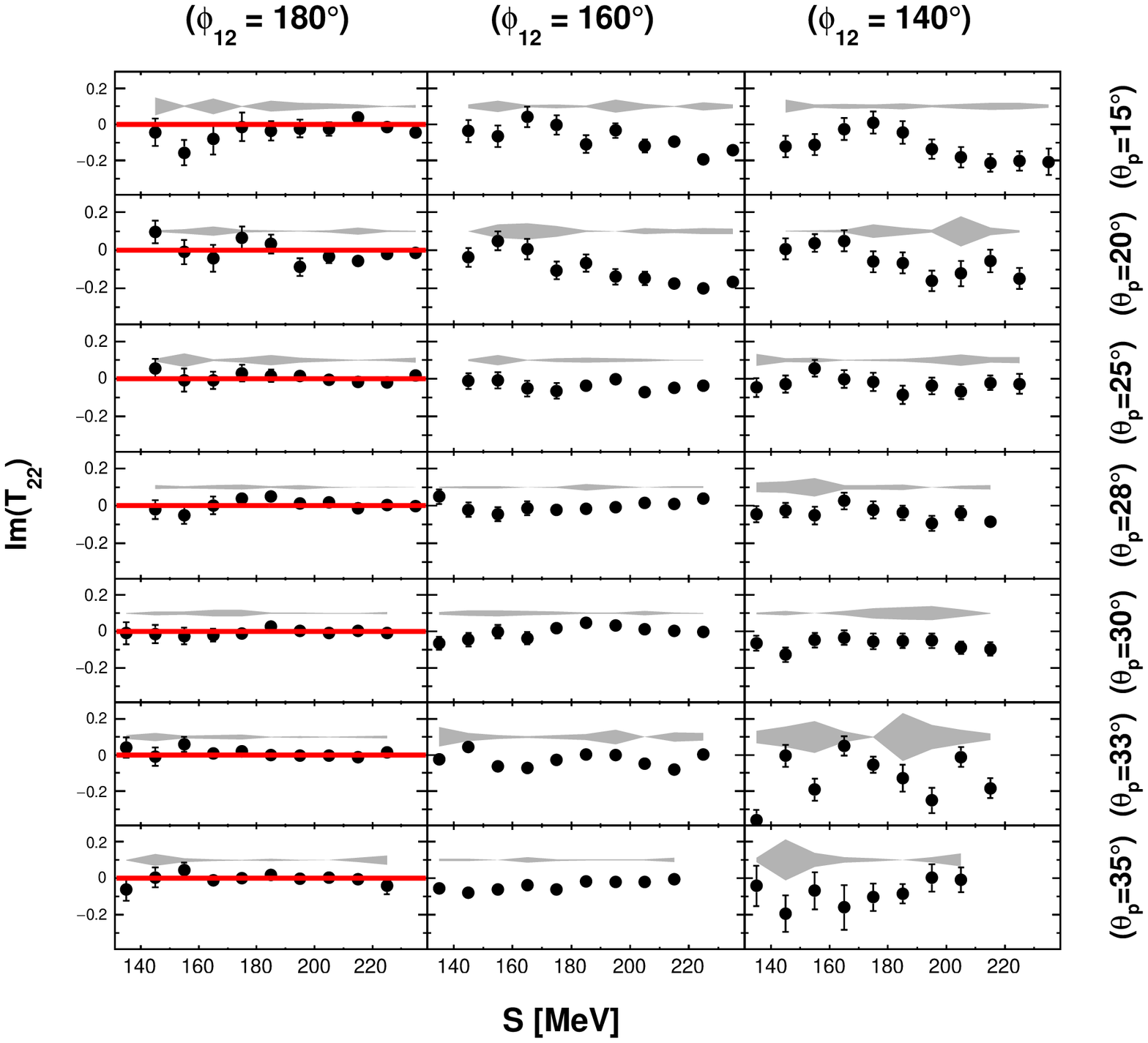}}
\vspace*{-0.4cm}
\caption{Same as Fig.~\ref{ImT22_15} except for $\theta_d = 28^\circ$.}
\label{ImT22_28}
\end{figure*}

\begin{thebibliography}{} 
%
\bibitem{Yukaw}        H. Yukawa,     \rm   Proc. Phys. Math. Soc. Jpn.  \textbf{17},   48      (1935).
\bibitem{Epel1}     E. Epelbaum \it{et al.},     \rm   Nucl. Phys. A     \textbf{671}, 295  (2000).
\bibitem{Epel2}     E. Epelbaum \it{et al.},     \rm   Phys. Rev. C     \textbf{65}, 044001  (2002).
\bibitem{wir}          R. B. Wiringa   \it{et al.},    \rm   Phys. Rev. C     \textbf{29},   1207         (1984).
\bibitem{sakai00}      H. Sakai         \it{et al.},     \rm   Phys. Rev. Lett. \textbf{84},   5288         (2000).
\bibitem{wit0}         H. Wita\l a     \it{et al.},        \rm Few Body Syst. \textbf{49} 61   (2010).
\bibitem{wita9}         H. Wita\l a     \it{et al.},        \rm Phys. Rev. C \textbf{77} 034004   (2008).
\bibitem{prima}        H. Primakoff  \it{et al.},   \rm   Phys. Rev.       \textbf{55},   1218         (1939). 
\bibitem{piep}         S. Pieper     \it{et al.},   \rm   Phys. Rev. C     \textbf{64}, 1  (2001).
\bibitem{Epel}         E. Epelbaum,   \rm   Prog. Part. Nucl. Phys.     \textbf{57}, 654  (2006).
\bibitem{wita}         H. Wita\l a      \it{et al.}, 	 \rm   Phys. Rev. Lett. \textbf{81},    11  (1998).
\bibitem{nasser1}       N. Kalantar-Nayestanaki  \it{et al.},   \rm    Rep. Prog. Phys.  \textbf{75}, 016301  (2012).
\bibitem{kars01}       K. Ermisch      \it{et al.},      \rm   Phys. Rev. Lett. \textbf{86},   5862         (2001).
\bibitem{kars03}       K. Ermisch      \it{et al.},      \rm   Phys. Rev. C     \textbf{68},   051001(R)    (2003).
\bibitem{kars05}       K. Ermisch      \it{et al.},      \rm   Phys. Rev. C     \textbf{71},   064004       (2005).
\bibitem{kimiko05}     K. Sekiguchi     \it{et al.},     \rm   Phys. Rev. Lett. \textbf{95},   162301       (2005). 
\bibitem{postma}       H. Postma and R. Wilson           \rm   Phys. Rev.       \textbf{121},  1129         (1961).
\bibitem{hamid}        H. Amir-Ahmadi   \it{et al.},     \rm   Phys. Rev. C     \textbf{75},   041001(R)    (2007).         
\bibitem{kurodo}       K. Kuroda        \it{et al.},     \rm   Nucl. Phys.       \textbf{88},   33           (1966). 
\bibitem{mermod}       P. Mermod        \it{et al.},     \rm   Phys. Rev. C     \textbf{72},   061001(R)    (2005).
\bibitem{Igo}          G. Igo           \it{et al.},     \rm   Nucl. Phys. A    \textbf{195},  33           (1972).
\bibitem{ald}          R. E. Adelberger      \it{et al.},   \rm   Phys. Rev. D     \textbf{5},    2139         (1972).
\bibitem{Hos}          H. Mardanpour    \it{et al.},     \rm   Eur. Phys. J. A  \textbf{31},   383          (2007).
\bibitem{Ela07}        E. Stephan       \it{et al.},     \rm   Phys. Rev. C     \textbf{76},   057001       (2007).
\bibitem{shimi}        H. Shimizu       \it{et al.},     \rm   Nucl. Phys. A    \textbf{382},  242          (1982).
\bibitem{hatan}        K. Hatanaka      \it{et al.},     \rm   Eur . Phys. J. A \textbf{18},   293          (2003).
\bibitem{IUCF}         E. J. Stephenson \it{et al.},     \rm   Phys. Rev. C     \textbf{60},   061001       (1999).
\bibitem{Ahmad1}       A. Ramazani-Moghaddam-Arani \it{et al.},     \rm   Phys. Rev. C     \textbf{78},   014006 (2008).
\bibitem{Ahmad2}       A. Ramazani-Moghaddam-Arani \it{et al.},     \rm   Few-Body Syst   \textbf{44},   27 (2008).
\bibitem{st1}          St. Kistryn   \it{et al.},     \rm   Phys. Rev. C     \textbf{68},   054004          (2003).
\bibitem{st2}          St. Kistryn   \it{et al.},     \rm   Phys. Rev. C     \textbf{72},   044006          (2005).
\bibitem{st3}          St. Kistryn   \it{et al.},     \rm   Phys. Lett. B     \textbf{641},   23          (2006).

\bibitem{hos2}         H. Mardanpour  \it{et al.},     \rm   Phys. Lett. B     \textbf{687},   149         (2010).
\bibitem{steph}         E. Stephan  \it{et al.},     \rm   Phys. Rev. C     \textbf{82},   014003         (2010).
\bibitem{Ic4}       I.~Ciepa\l~\it{et al.}, 	 \rm  	Few-Body Syst 56, 665-690 (2015).
\bibitem{Adl}       P. Adlarson        \it{et al.},     \rm   Phys. Rev. C     \textbf{101},  044001   (2020).
\bibitem{Par}       W. Parol        \it{et al.},     \rm   arXiv:2004.02651   (2020).
\bibitem{Haj}        H. Tavakoli-Zaniani      \it{et al.},     \rm   Eur . Phys. J. A \textbf{56},   62   (2020).
\bibitem{Dad}        M. Mohammadi-Dadkan      \it{et al.},     \rm   Eur . Phys. J. A \textbf{56},   3   (2020).
\bibitem{Ahmad4}       A. Ramazani-Moghaddam-Arani \it{et al.},     \rm   Phys. Rev. C     \textbf{83},   024004 (2011).

\bibitem{nemoto}       S. Nemoto        \it{et al.},     \rm   Phys. Rev. C     \textbf{58},    2599         (1998).

\bibitem{phill}        T. W. Phillips    \it{et al.}, 	 \rm   Phys. Rev. C \textbf{22},   384         (1980).
\bibitem{vivi}         M. Viviani    \it{et al.}, 	 \rm   Phys. Rev. Lett. \textbf{86},   3739     (2001).
\bibitem{fish}         B. M. Fisher    \it{et al.}, 	 \rm   Phys. Rev. C \textbf{74},    034001         (2006).
\bibitem{bech}         V. Bechtold,    \it{et al.}, 	 \rm   Nucl. Phys. A \textbf{288},   189         (1977).
\bibitem{Aldr}         C. Alderliesten    \it{et al.}, 	 \rm   Phys. Rev. C \textbf{18},   2001         (1978).
\bibitem{Garc}         M. Garcon     \it{et al.}, 	 \rm   Nucl. Phys. A  \textbf{458},   287         (1986).
\bibitem{Micher}       A. M. Micherdzinska    \it{et al.}, 	 \rm   Phys. Rev. C \textbf{75},   054001    (2007).
\bibitem{Ic1}       I.~Ciepa\l~\it{et al.}, 	 \rm   Phys. Rev. C \textbf{99},   014620    (2019).
\bibitem{Ic2}       I.~Ciepa\l~\it{et al.}, 	 \rm   Phys. Rev. C \textbf{100}, 024003 (2019).
\bibitem{Ic3}       I.~Ciepa\l~\it{et al.}, 	 \rm  	Few-Body Syst 60: 44 (2019).
\bibitem{myp2}         R. Ramazani-Sharifabadi \it{et al.}, \rm IL NUOVO CIMENTO \textbf{42 C}  129 (2019).
\bibitem{Det1}        A. Deltuva and A.C. Fonseca, \rm   Phys. Rev. C \textbf{86},  011001(R) (2012).
\bibitem{Det2}        A. Deltuva and A.C. Fonseca, \rm   Phys. Rev. C \textbf{87},  054002(R) (2013).
\bibitem{Det3}        A. Deltuva and A.C. Fonseca, \rm   Phys. Rev. Lett. \textbf{113},  102502 (2014).
\bibitem{Det4}        A. Deltuva and A.C. Fonseca, \rm   Phys. Rev. C \textbf{90},  044002 (2014).
\bibitem{Det5}        R. Lazauskas, \rm   Phys. Rev. C \textbf{91},  041001(R) (2015).
\bibitem{Det6}        A. Deltuva and A.C. Fonseca,    \rm   Phys. Rev. C \textbf{91},  034001 (2015).
\bibitem{Det7}        A. Deltuva and A.C. Fonseca,    \it{et al.}, \rm   Phys. Lett. B \textbf{742},  285-289 (2015).
\bibitem{Det8}        A. Deltuva and A.C. Fonseca,    \it{et al.}, \rm   Phys. Rev. C \textbf{95},  024003 (2017).
\bibitem{Delt1}        A. Deltuva    \it{et al.}, 	 \rm   Phys. Rev. C \textbf{68},  024005 (2003).
\bibitem{Delt2}        A. Deltuva and A.C. Fonseca, 	 \rm   Phys. Rev. C \textbf{92},  024001 (2015).
\bibitem{Delt3}        A. Deltuva and A.C. Fonseca, 	 \rm   Phys. Rev. C \textbf{93},  044001 (2016).
\bibitem{myp1}         R. Ramazani-Sharifabadi \it{et al.},  \rm Eur. Phys. J. A \textbf{55}, 177 (2019).

\bibitem{Ahmad40}       A. Ramazani-Moghaddam-Arani \it{et al.},     \rm   EPJ Web of Conferences  \textbf{3}, 04012 (2010).
\bibitem{Fri}       L. Friedrich    \it{et al.}, 	 \rm   Polarized beams and polarized gas targets (World Scientific, Singapore) , p. 198  (1995).
\bibitem{Krem}     H. R. Kremers    \it{et al.}, 	Polarized Gas Targets and Polarized Beams,  \rm   AIP Conj Proc. \textbf{421},   p. 507    (1997).
\bibitem{nasser3}       N. Kalantar-Nayestanaki    \it{et al.},        \rm   Nucl. Instrum. and  Meth. in  Phys. Res.  A \textbf{417},  215 (1998).
\bibitem{mythesis}       R. Ramazani-Sharifabadi, Ph.D. thesis, University of Groningen, (2020).
\bibitem{Ohl}       G. G. Ohlsen,       \rm   Nucl. Instrum. Meth. \textbf{179},  283 (1981).
\bibitem{nasser}       N. Kalantar-Nayestanaki.  \it{et al.},   \rm    Nucl. Instrum. and Meth. in Phys. Res. A \textbf{444}, 591  (2001).
\bibitem{hos3}         H. Mardanpour, Ph.D. thesis, University of Groningen, (2008).
\bibitem{Ahmad3}       A. Ramazani-Moghaddam-Arani, Ph.D. thesis, University of Groningen, (2009).
\bibitem{BBSr}       C. D. Bailey, Ph.D. thesis, Indiana University, (2009).
\bibitem{lsp}          H. R. Kremers    \it{et al.}, 	 \rm   Nucl. Instrum. and Meth. in Phys. Res. A \textbf{516},  209 (2004).
\bibitem{ibp}          R. Bieber        \it{et al.}, 	 \rm   Nucl. Instrum. and Meth. in Phys. Res. A \textbf{457},  12 (2001).
\bibitem{ge3}       J. Allison   \it{et al.},  \rm IEEE Transactions on Nuclear Science \textbf{53}, 1 (2006).
\end{thebibliography}
\end{document}